\documentstyle[12pt,aasms4]{article}

\lefthead{Gonzalez, Laws, and Ward}
\righthead{The Galactic Habitable Zone}
\slugcomment{accepted by Icarus}
\begin{document}

\title{The Galactic Habitable Zone I. Galactic Chemical Evolution}
\author{Guillermo Gonzalez, Donald Brownlee}
\affil{Astronomy Department, University of Washington, P.O. Box 351580, 
Seattle, WA 98195 USA}
\author{Peter Ward}
\affil{Department of Geological Sciences, University of Washington, P.O. 
Box 351310, Seattle, WA 98195 USA}
\authoremail{gonzalez@astro.washington.edu}

\begin{abstract}

We propose the concept of a "Galactic Habitable Zone" (GHZ).  Analogous to 
the Circumstellar Habitable Zone (CHZ), the GHZ is that region in the Milky 
Way where an Earth-like planet can retain liquid water on its surface and 
provide a long-term habitat for animal-like aerobic life.  In this paper we 
examine the dependence of the GHZ on Galactic chemical evolution.  The single 
most important factor is likely the dependence of terrestrial planet mass on 
the metallicity of its birth cloud.  We estimate, very approximately, that a 
metallicity at least half that of the Sun is required to build a habitable 
terrestrial planet.  The mass of a terrestrial planet has important 
consequences for interior heat loss, volatile inventory, and loss of 
atmosphere.  A key issue is the production of planets that sustain plate 
tectonics, a critical recycling process that provides feedback to stabilize 
atmospheric temperatures on planets with oceans and atmospheres.  Due to the 
more recent decline from the early intense star formation activity in 
the Milky Way, the concentration in the interstellar medium of the 
geophysically important radioisotopes, $^{40}$K, $^{235, 238}$U, $^{232}$Th, 
has been declining relative to Fe, an abundant element in the Earth.  Also 
likely important are the relative abundances of Si and Mg to Fe, which 
affects the mass of the core relative to the mantle in a terrestrial planet.  
All these elements and isotopes vary with time and location in the Milky Way; 
thus, planetary systems forming in other locations and times in the Milky Way 
with the same metallicity as the Sun will not necessarily form habitable 
Earth-like planets.  As a result of the radial Galactic metallicity gradient, 
the outer limit of the GHZ is set primarily by the minimum required 
metallicity to build large terrestrial planets.  Regions of the Milky Way 
least likely to contain Earth-mass planets are the halo (including globular 
clusters), the thick disk, and the outer thin disk.  The bulge 
should contain Earth-mass planets, but stars in it have a mix of elements 
different from the Sun's.  The existence of a luminosity-metallicity 
correlation among galaxies of all types means that many galaxies are 
too metal-poor to contain Earth-mass planets.  Based on the observed 
luminosity function of nearby galaxies in the visual passband, we estimate 
that: 1) the Milky Way is among the 1.3\% most luminous (and hence most 
metal-rich) galaxies, and 2) about 23\% of stars in a typical ensemble of 
galaxies are more metal-rich than the average star in the Milky Way.  The GHZ 
zone concept can be easily extrapolated to the universe as a whole, 
especially with regard to the changing star formation rate and its effect on 
metallicity and abundances of the long-lived radioisotopes.

\end{abstract}


\section{INTRODUCTION}

While its definition has varied somewhat over the last four decades, the 
Circumstellar Habitable Zone (CHZ) has generally been defined to be that 
region around a star where liquid water can exist on the surface of a 
terrestrial (i.e., Earth-like) planet for an extended period of time (Huang 
1959; Shklovsky and Sagan 1966; Hart 1979).  Estimates of its size have varied 
as climate models have been refined (Kasting et al. 1993; Franck et al. 
2000b), but the present models do have some shortcomings.  The most serious 
one is the adoption of the present Earth as the standard when applying the 
CHZ to other locations in the universe or other epochs in Earth's history.  
For example, the width of the CHZ is influenced by chemical weathering, but 
its efficiency is greatly increased by the presence of vascular plants (see 
Moulton and Berner 1998).  Therefore, application of extant CHZ models to 
other planetary systems assumes that advanced plant life is already present.  
This inherent inconsistency notwithstanding, the CHZ is still a useful concept 
around which the presence of advanced life in the universe can be considered 
in a quantitative way.

Here we propose that, in an analogous fashion, there exists within galaxies a 
region favorable to the development and long-term maintenance of complex life 
comparable to terrestrial animals and complex plants.  In the Milky Way this 
region forms an annulus in the disk, the boundaries of which are set by 
several Galactic-scale astrophysical processes that are likely to relate to 
habitability on terrestrial planets.  Its location is variable with time, 
due, for example, to the evolution of the abundances of heavy elements 
(including the long-lived radioactive isotopes) in the interstellar medium.  
The inner limit of this newly defined Galactic Habitable Zone (GHZ), is set by 
high energy events such as supernova and gamma ray bursts, higher metallicity 
leading to orbital decay of planets, and gravitational perturbations of 
Oort cloud comets sufficient to cause frequent comet impacts (to be discussed 
in a future paper).  These events cause mass extinctions and perturb complex 
life.  The outer limit of the GHZ, the subject of this paper, is set by 
Galactic chemical evolution, and, in particular, the radial disk metallicity 
gradient.  A certain minimum abundance of heavy elements is needed to fully 
assemble Earth-size planets in the CHZ.  In addition, we argue that certain 
elements must be present with the right mix.  The boundaries of the GHZ are 
not rigid, well definable limits, but are rather probabilistically defined.

Ours is not the first discussion of Galactic-scale constraints on 
habitability.  Trimble (1997a,b) considered, in a general way, the 
composition of the Sun in the broader context of Galactic chemical evolution 
and how heavy element abundances in the interstellar medium constrains the 
timing and location of habitable planets.  Clarke (1981) discussed the 
possible limitations on the habitability of a galaxy undergoing a Seyfert-like 
outburst.  Several astronomers have been arguing since the early 1980's 
(e.g., Balazs 1988; Marochnik 1984) that the placement of the Sun's galactic 
orbit very near the corotation circle is an important requirement for 
habitability.  Finally, many papers have been published since the early 1970's 
about the possible damaging effects of a nearby supernova (e.g., Brakenridge 
1981; Ellis and Schramm, 1995).  Each of these previous studies focused on 
only one type of Galactic-scale constraint.  Tucker (1981) was one of the few 
to consider habitability in the Galaxy within a broader framework, but his 
treatment was very superficial and is now seriously outdated.
Much of our motivation for investigating the possible link between 
Galactic-scale astrophysical processes and life on a terrestrial planet 
derives from the apparently anomalous values of several of the Sun's 
parameters, which Gonzalez (1999a,b) has interpreted within the framework of 
the Weak Anthropic Principle.  For example, it seems an odd coincidence that 
we should happen to be living around a star with a space velocity relative to 
the Local Standard of Rest ($\nu_{\rm LSR}$) smaller than most other nearby 
stars.  The oddity of this situation can be removed if it can be shown that 
habitability favors a small value of $\nu_{\rm LSR}$.  While this point was 
already addressed by Gonzalez (1999a,b), our present task is to bring the 
discussion within the framework of the GHZ.  Our motivation also derives from 
much recent research showing that mass extinctions on Earth have severely 
affected the course of biotic evolution, and seemingly pose a threat to the 
survival of complex life on the surface of any planet (Ward and Brownlee 2000).

Our purpose herein is to present a unified, though not necessarily complete, 
treatment of the GHZ, including astrophysical processes over a wide range of 
space and time scales.  In this first paper in a series we will address the 
constraints imposed on the GHZ by Galactic chemical evolution, focusing on 
the production history of biologically important elements.  In future papers 
we will discuss threats from transient radiation events and asteroid and comet 
impacts and give quantitative estimates of the size and time evolution of the 
GHZ in the Milky Way and in other galaxies.  In the following we begin with a 
presentation of our working definition of a habitable planet and a discussion 
of the basic geophysical requirements, followed by a discussion of the 
astrophysical processes that provide the basic planetary building blocks, 
compare the Milky Way habitability to other nearby galaxies, and end with a 
summary of the GHZ concept.

\section{DEFINING A HABITABLE PLANET}

Throughout the following we adopt the Earth as the reference habitable 
terrestrial planet, which we do for two simple reasons: 1) it is the only 
example we have, and 2) comparative planetology in the Solar System indicates 
that the Earth's habitability may be near optimal (especially for complex 
life).  However, this second assumption may not be true for every parameter; 
for instance, the Earth may not be not have an optimal impact rate (impacts 
can have both positive and negative consequences).  We employ the term 
habitability to refer to environmental requirements suitable for Earth-like 
animal-like aerobic life, not the broader range of conditions that might 
support microbial life.  We have identified three necessary (though not 
sufficient) requirements for such a habitable terrestrial planet: 1) an ocean 
and some dry land, 2) moderately high O$_{\rm 2}$ abundance, and 3) 
long-term climate stability.  The moderately high O$_{\rm 2}$ (and low 
CO$_{\rm 2}$) abundance is a necessity for large mobile life on 
physiological grounds and also for building an ozone 
shield (McKay 1996).  An ocean is required primarily for temperature 
regulation via the operation of a water cycle on a global scale, plate 
tectonics, and chemical weathering (on land).  A completely ocean covered 
planet is excluded from consideration, because it eliminates solid 
surface-atmosphere interactions and limits the diversity of possible life 
(life on such a ``waterworld'' would likely be limited to that typical of 
deep sea thermal vents on the present Earth, if even that much).  Long-term 
climate stability brings in many astrophysical and geophysical constraints, 
such as: stellar evolution, comet and asteroid impact rate, the presence of a 
large natural satellite, and a long-term planetary heat source to drive plate 
tectonics.  In the present study we address those steps in the formation of 
such a habitable planet that are linked to the broader topic of Galactic 
chemical evolution.  Implicit in our study is the assumption that deviations 
from Earth-like parameters lead to less habitable conditions.  Full 
verification of this assumption may be possible in the future as models 
integrating planetary dynamics, geophysics, climate, and biology are further 
refined.

As an aside, we note that a large natural satellite orbiting a gas giant 
planet has been suggested as a possible alternative habitat to an Earth-like 
planet (e.g., Williams et al. 1997).  However, such an environment is very 
likely to be less habitable for the following reasons: 1) comet collisions 
will be more frequent due to the strong gravity of the gas giant host, 
especially via captures into temporary orbits (also more frequent comet 
breakups will contribute to an increased collision probability), 2) the 
particle radiation levels are higher in the vicinity of a gas giant with a 
strong magnetic field like Jupiter (but some gas giants, like Saturn, have 
weaker fields), and 3) spin-orbit tidal locking will occur on a relatively 
short timescale.  The first point was illustrated by the capture and 
subsequent breakup of comet Shoemaker-Levy 9 in 1994; it is also illustrated 
by the discovery of crater chains on Callisto and Ganymede (Schenk et al. 
1996 examined 116 craters in 11 crater chains).  Of course, not every 
planetary system will have the same comet flux as ours, due to different 
formative histories and giant planet configurations.  The second point leads 
to a more rapid loss of an atmosphere for a satellite without a strong 
magnetic field and is a threat to surface life.  The third point leads to 
greater day/night temperature swings.  For environments similar to the 
Galilean satellite system, tidally locked orbits can be beneficial in that 
they generate internal heat through tidal stressing, but orbital changes on 
timescales of 10$^8$ to 10$^9$ years prevents tides from being a continuously 
available source of heat (Greenberg et al. 2000).  The problems noted above 
are somewhat mitigated if the satellite orbits far from its host planet, but 
in such a configuration the insolation from the parent star becomes more 
variable.  Finally, it is not clear if a giant planet can end up at $\sim$0.5 -
5 AU from its host (G - K spectral type) star in an orbit as nearly circular 
as the Earth's (the present eccentricity of the Earth's orbit is 0.017).  To 
date, all the giant planets found around other solar type stars beyond 
$\sim$0.15 A.U. have much more eccentric orbits than does the Earth, 
except HD 27442, which has an eccentricity of 0.025 (Butler et al. 2001); 
even Jupiter has an eccentricity of 0.048.  Therefore, since they are inferior 
habitats for complex life, we will not consider natural satellites further.

\subsection{Planet Mass}

It is likely that planet mass is the single most important factor in building 
a habitable terrestrial planet.  We make the simplifying assumption that a 
terrestrial planet's mass is determined primarily by the local surface 
density, $\sigma$, of the protoplanetary disk from which it forms.  Following 
the equations describing planet formation in Lissauer (1995), we assume that 
terrestrial planet mass scales with $\sigma^{1.5}$.  This dependence is 
determined both by the surface density of solids in a feeding zone and the 
narrower width of a feeding zone of a smaller planet.  The complexities of 
planet formation may cause significant deviations from this simple functional 
dependence, but we believe that this relationship will provide a reasonably 
close estimate of typical planetary mass.  We also assume that: 1) the surface 
density at a radial location in a protoplanetary disk is directly proportional 
to the heavy element abundance of the interstellar cloud out of which a 
planetary system condenses, and 2) the total mass of terrestrial planets 
that form in a given planetary system is scalable from the Solar System in 
proportion to the relative heavy element abundances of their parent stars.  
Until the results of simulations which implicitly include the composition of 
the birth cloud are made available, we will have to make do with these 
assumptions.

Present observations of extra-solar-system\footnote{Although the preferred 
term in the literature is ``extrasolar'', we will employ the more 
grammatically correct ``extra-solar-system'' in the present paper.} planets 
can be of some help to us in determining some astrophysical constraints on 
habitability.  The most important findings to date are: 1) the high mean 
metallicity\footnote{Note, throughout this paper, we employ the astronomical 
definition of metals, i.e., element heavier than He.} of stars with planets 
compared to the general field stars, 2) the very short orbital periods of 
some planets, and 3) the high eccentricities of planets with orbital periods 
greater than about two weeks (Butler et al. 2000).  Combining the first point 
above with the lack of detection of giant planets in the globular cluster 47 
Tucanae (Gilliland et al. 2000) implies that a minimum metallicity near 40\% 
solar is required to build giant planets.  Among the stars with close-in 
planets, the metallicities are particularly high.  This could be due to disk 
material falling onto the star and polluting its atmosphere, and/or a 
metallicity dependence on giant planet formation, and/or to a metallicity 
dependence of planet migration (see Gonzalez et al. 2001 for additional 
discussion on these points).  If the latter is the case, then planet 
migration may occur more frequently in metal-rich systems, leading to the 
disruption of the orbits of any terrestrial planets in the habitable zone.  
Similarly, if the high eccentricities of the orbits of the giant planets 
observed around other stars are due to planet-planet interactions (see 
Weidenschilling and Marzari 1996), this phenomenon, too, is more likely to 
occur in metal-rich systems.  The metallicity dependencies of these phenomena 
are not yet well-constrained, so we will not include them in the 
present study (for an attempt to treat these phenomena quantitatively, see 
Lineweaver 2000).

It is useful to consider the sensitivity of habitability to changes in planet 
mass.  Lewis (1998) notes three important differences between the Earth and a 
smaller or larger sibling: 1) heat flow and associated geophysical processes, 
2) volatile inventory, and 3) atmosphere loss rate.  He estimates that 
an Earth-like planet identical in composition to the Earth with the same 
orbit but with one-tenth the mass (about half the radius) would have a 
lithosphere over twice as thick, suffer much more rapid loss of atmosphere 
from impact explosive blowoff and dissociative recombination in the exosphere, 
and have an ocean only 20\% as deep.  At the opposite extreme, a planet with 
12 times the mass (about twice the radius) would have a much thinner 
lithosphere, suffer no significant loss of its atmosphere, and have an 
average ocean depth over three times as great as the Earth's (with no dry 
land).  These estimates assume that the volatile reservoir is primarily the 
result of outgassing and that its efficiency is proportional to the size of 
the planet.

There are a couple of other size-dependent factors, which Lewis does not 
discuss.  One concerns impact probability.  It is proportional to the square 
of a planet's the radius (plus a little more owing to gravitational 
focusing).  Furthermore, the increased impact rate would apply to all size 
scales of impactors; very rare, very large impacts would occur more often on 
a larger planet.  However, the impact energy on a larger planet is diluted 
over an area proportional to the square of its radius.  A proper evaluation 
of the net dependence of impacts on planetary size will require careful 
consideration of the details of the relevant processes (such as the mass 
function of the impactor population).  Another size-dependent factor is the 
time to buildup O$_{2}$ in the atmosphere (McKay 1996).  It should take more 
time to oxygenate the atmosphere of a larger terrestrial planet, due to its 
larger inventory of reductants.  However, a complete analysis of this factor 
has yet to be published.  For the purposes of the present discussion, we 
follow Lewis (1998) and require that a terrestrial planet with the same orbit 
and formation history of the Earth have a mass between one third and three 
Earth masses to be habitable, though it is our opinion that this range is 
probably too broad.

We note that the above discussion must be placed in the broader context of 
the terrestrial planet environment.  In particular, a cloud that is 
sufficiently metal-rich to form giant Earth siblings will also likely form 
smaller terrestrial planets (cf. Wetherill 1996).  Therefore, the problems 
noted above for a very large terrestrial planet in a given system would still 
not prevent that system from having a terrestrial planet in the required mass 
range.  The same cannot be said of a metal-poor cloud, which forms only 
small terrestrial planets.  Given this, then, habitability for a given system 
cannot be excluded solely on the basis that it forms one or more massive 
Earths, but it might be excluded on other grounds related to high initial 
metallicity (as noted earlier).

\subsection{Plate Tectonics}

Of the geophysical processes that affect habitability, plate tectonics is 
probably the most critical, because it plays a central role in maintaining 
earth's long-term climate stability (Kasting et al. 1993; Franck et al. 1999, 
2000a).  Subduction of carbonates, and their ultimate thermal decomposition, 
provides the volcanic CO$_{\rm 2}$ source that drives the CO$_{\rm 2}$-rock 
cycle.  The continuing CO$_{\rm 2}$ supply from subducted sediments combined 
with the temperature dependent removal of CO$_{2}$ by chemical weathering 
provides negative feedback that opposes major global temperature variations. 
Without plate tectonics it is unlikely the earth or similar planets could 
maintain habitable surface environments for animal life for long periods of 
time (Franck et al. 1999, 2000a).

Unfortunately, the requirements for the origin and maintenance of plate 
tectonics are not well known.  On earth the process produces spreading 
center ridges, trenches, subduction zones and linear mountain ranges.  From 
the lack of similar features on other bodies it is clear this process is not 
currently operating on any other solar system body.  Mars and Venus display a 
rigid lid, and lack plate tectonics at the present (Tackley 2000).  Mars is 
apparently too small.  Venus, a near twin of Earth, does not have plate 
tectonics probably because of its lack of water, which lowers the melting 
points of magmas and promotes general ductility required for subduction 
(Karato and Jung 1998).  A necessary, though certainly not sufficient, 
requirement for plate tectonics is a long-term supply of heat.  Radioactive 
decays of $^{235,238}$U, $^{232}$Th, and $^{40}$K are considered to be the 
primary sources of heat in the Earth's interior.  Very early in the Earth's 
history there were other sources of interior heat: short-lived radioisotopes, 
heat of formation (from accretion, including the lunar-forming impact), and 
formation of the core.  However, after the first 1.5 Gyrs, the initial 
thermal state of the Earth was erased due to a re-adjustment effect in 
mantle convection (see Franck 1998).  Therefore, following an early 
settling-down period, the heat flow through a terrestrial planet's mantle is 
largely determined by the abundances of long-lived radioactive isotopes in 
its interior.  However, the amount of volatile outgassing does depend rather 
sensitively on the initial thermal state (Franck 1998); hence, the amount of 
outgassing will be dependent on the details of the formation process.

\subsection{CHZ Limits}

The dimensions of the Circumstellar Habitable Zone (CHZ) in the Solar System 
are defined by more than just the flux of stellar radiation.  There are 
several other parameters that should be included in the definition if it is to 
be generalized to any location and time in the Milky Way.  The relevant 
parameters are: mass, composition, orbit, and type of parent star.  Taken 
together, the possible parameter space covered by these quantities is very 
large, and it is not possible to set separate constraints on them.  For 
example, the inner edge of the CHZ for a terrestrial planet smaller than the 
Earth would be farther from the Sun than the CHZ of the Earth due to the more 
rapid loss of its atmosphere.  However, there are a few simplifying 
assumptions we can make to narrow it significantly.  One is the restriction of 
the orbit to small eccentricity in order to prevent large temperature swings.  
Another one is the restriction to main sequence stars similar in spectral type 
to the Sun (early G), which finds support from the Weak Anthropic Principle 
(Gonzalez 1999b) and theoretical arguments concerning stellar lifetimes and 
tidal locking and observations of flare activity in low mass dwarfs.

There are at least three other phenomena/processes that are presently not 
included in studies of the CHZ, such as those cited in the Introduction.  One 
is the variation of the impact threat with position, due to the increase in 
encounter velocities with decreasing distance from the parent star.  The 
impact energy of an asteroid or comet is proportional to the square of the 
relative velocity between the impactor and the planet.  Another factor is the 
composition gradient of the early solar nebula.  Lewis (1997) describes 
quantitatively the condensation sequence of elements as a function of distance 
from the central star (described in greater detail below).  A third factor 
concerns the details of planet formation.  For example, the radial distribution 
of terrestrial planets is stochastic and is relatively insensitive to stellar 
mass according to simulations of the late stages of planet formation carried 
out by Wetherill (1996).  Taken together, these additional factors more 
severely constrain the dimensions of the CHZ.  However, a consideration of the 
complex relationships among the various factors defining the CHZ are beyond 
the scope of the present work.  We, instead, focus on a subset of these, which 
we will explore within the larger context of the Milky Way.

\subsection{Composition}

In addition to the overall metallicity (and hence mass) there are 
compositional factors that influence habitability.  A comparison between 
Earth and Venus illustrates that planet mass is not the only factor that 
determines a planet's geophysics.  Although they are nearly the same mass, 
Earth and Venus have very different geology.  Most of these differences 
probably date back to the different origins of Earth and Venus.  Lewis (1997) 
notes that the composition gradient of the early solar nebula inferred 
from the expected radial condensation sequence of the elements will result in 
planetary embryos of varying composition.  The early stages of planet 
formation are likely to draw from a relatively narrow range in distance from 
the central star, but latter accretion is likely to be from a relatively 
larger range (Wetherill 1996).  This implies that the properties of a 
terrestrial planet's core and mantle depend rather sensitively on its place of 
formation.  For example, more FeS should have condensed in the proto-Earth 
relative to the proto-Venus, and this would have lead to a significant 
reduction in the melting point of the Earth's core.  The relative abundance of 
$^{40}$K, an important source of long-term heat, would also have been 
variable; given its relatively high volatility, less of it would have been 
incorporated by protoplanets inside the Earth's orbit.  The CHZ definition, 
then, must include not only the dependence of solar insolation but also of 
bulk composition on radial distance.  The detailed dependencies of planetary 
composition on place of formation are beyond the scope of this paper, but it 
is a factor that must be kept in mind when we discuss Galactic chemical 
evolution.

The most abundant element in the Earth by number is O, while the most abundant 
element by mass is Fe (Kargel and Lewis 1993).  The next most abundant 
terrestrial elements are Mg and Si; the cosmically abundant elements, H, He, 
C, N, are present only in trace amounts (Kargel and Lewis 1993).  The O 
abundance can be considered as a free parameter, since it forms oxides with 
Mg, Si, and Fe.  Another factor likely to be relevant to the habitability of 
a terrestrial planet is the abundance of Mg and Si relative to Fe, which will 
determine the mass of the core relative to the mantle.  The condensation 
temperatures of Mg, Si, and Fe are similar\footnote{Lodders and Fegley (1998) 
list the condensation temperature of Mg, Si, and Fe as 1340, 1340, and 1337 K, respectively.}, so they should condense in ratios similar to that of their 
host star and terrestrial planets should 
experience relatively little differential fractionation.  In the Milky Way's 
disk, stars with low Fe/H have relatively larger Mg/Fe and Si/Fe ratios (see 
Section 3.1.1).  These ratios can differ by up to 250 percent from the solar 
values and may have major influence on the interior workings of a terrestrial 
planet, such as the efficiency of core formation, the nature of 
solid-state convention, and its mineralogy.

Gaidos (2000) has noted the likely significance of the C/O ratio in the 
formation of planets.  The C/O ratio determines how much oxygen in an 
accretion disk is in CO and how much is in water.  Water ice was an abundant 
condensable substance in the outer regions of our Solar System's early history 
and was an important ingredient in the formation of the embryos of the giant 
planets.  As discussed above, it is also an essential ingredient in forming a 
habitable terrestrial planet.  The bulk C/O ratio in a planet will 
be the result of a variety of fractionation and volatile delivery processes, 
so one cannot just equate it to the initial C/O ratio in a protoplanetary 
nebula.  The relevance of the C/O ratio for the formation of habitable 
terrestrial planets is probably less direct, affecting on the characteristics 
of the giant planets and the population of cometary bodies.  However, Gonzalez 
et al. (2001) do not detect a significant difference in the C/O ratio between 
the parent stars of extra-solar-system planets and the general field star 
population (the previously reported low C/O ratio for the Sun compared to 
field stars has proven to be spurious, due to an unrecognized systematic error 
in C abundance estimates from one of the cited studies).  Therefore, support 
for the importance of the C/O ratio to planet formation currently comes only
from theoretical arguments.

\subsection{Basic Requirements}

In summary, sufficient metals must be available in the interstellar medium 
to build a habitable terrestrial planet.  Following Lewis (1998), we assume 
its mass should be between approximately one third and three Earth masses.  In 
addition, the relative proportions of Si, Mg, and Fe should be similar to 
those in the Earth, and there are minimum required abundances of the 
long-lived radioactive isotopes, $^{235,238}$U, $^{232}$Th, and $^{40}$K.

The abundance of $^{40}$K in a terrestrial planet's crust may be relevant 
also to the origin of life.  Draganic et al. (1991) noted that the ionizing 
radiation produced from the decay of $^{40}$K in the oceans of the early Earth 
would have generated O$_{2}$ and H$_{2}$O$_{2}$ via radiolysis of water.  The 
early stages of the extant chemical evolution of life scenarios require 
reducing conditions, so a planet with too much $^{40}$K mixed in its oceans 
may delay or prevent the appearance of life.

\section{GALACTIC EVOLUTION AND SPATIAL DISTRIBUTION OF THE ELEMENTS}
\subsection{The Stable Elements}

Since the formation of the Milky Way about 12 to 15 Gyr ago, the metals have 
increased in abundance in the interstellar medium (ISM) near the Sun from near 
zero to about 2\% by mass.  To first order, the cosmic abundances of the 
elements heavier than Boron scale with Fe, which has many easily measured 
absorption lines in the spectra of Sun-like stars.  Therefore, what is 
sometimes termed the ``metallicity'' of a star is often its Fe abundance -- in 
the following we will employ the terms [Fe/H]\footnote{[X/Fe] $\equiv \log 
(N_{\rm X}/N_{\rm Y})_{\rm Star} - \log (N_{\rm X}/N_{\rm Y})_{\rm Sun}$, 
where the variable N corresponds to the number density abundance of element X 
and Y. Hence, a star with [Fe/H] $= +0.5$ has 3 times the Fe/H abundance 
ratio of the Sun.} and metallicity interchangeably.  Observational and 
theoretical knowledge of the evolution of the chemical element abundances 
in the Milky Way has improved greatly over the 
last 30 years.  The overall abundance trends of most elements are now well 
established from spectroscopic observations of stars in the solar neighborhood 
(e.g., Gratton et al. 2000; Chen et al. 2000; reviewed by McWilliam 1997).  In 
the following we will consider the evolution of both the mean Fe abundance and 
abundance ratios, as X/Fe, and discuss their relation to habitability as 
defined in Section 2.

The chemical evolution of the ISM is governed primarily by the rate of infall 
of unprocessed gas and processing of ISM matter by massive and intermediate 
mass stars (Timmes et al. 1995; Pagel 1997; Portinari et al. 1998; Samland 
1998).  The primary specific sources of the metals are 
supernovae\footnote{Except C and N, for which a substantial contribution is 
from intermediate mass stars.}, both those deriving their energy from nuclear 
explosions of white dwarfs (type Ia) and those deriving their energy from core collapse (type Ib,c; type II).\footnote{Hereafter, SNe Ib,c II will be 
grouped together as SNe II.} SNe II are the primary source of O, 
the $\alpha$\footnote{The $\alpha$ elements typically measured in stellar 
spectra are Mg, Si, Ca, and Ti.}, and the $r$-process elements, and SNe Ia are 
the primary source of the Fe-peak (i.e., Fe, Co, Ni) elements (Timmes et al. 
1995; Portinari et al. 1998; Samland 1998).  In the simplest terms, the 
evolution of abundance ratios can be understood in terms of the changing 
relative contributions from these two basic SN types, which can be due to 
changes in the initial mass function and in the star formation rate.  A 
particularly important constraint on Galactic chemical evolution models is the 
observed decline in the O/Fe ratio with increasing Fe/H among nearby stars 
(e.g., Edvardsson et al. 1993).  This is thought to result from the rise in 
the frequency of SNe Ia with time relative to that of SNe II (Mathews et al. 
1992; see Fig. 39 of Timmes et al. 1995, Fig. 12 of Portinari et al. 1998, and 
Fig. 1 of Samland 1998).

When discussing chemical evolution, it is convenient to divide up the Milky 
Way into four components: thin disk, thick disk, bulge, and halo.  While they 
partially overlap in space, these components can be relatively well separated 
with the addition of data on stellar kinematics.  The thin disk, of which the 
Sun is a member, includes significant ongoing star formation activity, and its 
kinematics is more nearly purely rotational than the other components; stars 
extend about 180 pc on either side of the midplane.  The thick disk is 
basically a fatter version of the thin disk and contains stars that are more 
metal-poor than it and extend about 600 pc on either side of the midplane.  
The bulge is centered at the center-of-mass of the Milky 
Way.\footnote{Here we are assuming that the thin and thick disks each has a 
central hole.  Therefore, given this assumption, only bulge stars occupy 
the central region of the Milky Way.} Its constituent stars display a large 
range in metallicity.  In the Milky Way the bulge experienced an early rapid 
rate of star formation, which has since slowed, but continues to the present 
day.  The halo is a spherical distribution of individual stars and globular 
clusters extending to about 100 kpc.  It is the oldest, most metal-poor 
component.  We discuss each of these components separately below.

\subsubsection{The Thin Disk}

The overall metallicity of the ISM in the thin disk has been increasing 
steadily (but episodically on short timescales; Rocha-Pinto et al. 2000).  
Gonzalez (1999a) derived the following equation from a sample of nearby F and 
G dwarfs with well-determined physical parameters, relating [Fe/H] to age and 
to mean Galactocentric distance, R$_{m}$:

\begin{equation}
[Fe/H] = (-0.01\pm0.02) - (0.07\pm0.01) (R_{\rm m} - R_{\rm 0})
- (0.035\pm0.005)\tau
\end{equation}

where $\tau$ is the age in Gyr and R$_{0}$ is the present Galactocentric 
distance of the Sun (we adopt R$_{0}$ = 8.5 kpc in the present study).  Since 
this equation is determined from a sample of nearby stars, it is not 
applicable to the halo or to the bulge of the Milky Way, which have had 
significantly different histories from the local disk.  The assumption of a 
linear dependence of [Fe/H] on age is justifiable for stars no older than 
about 80\% of the age of the Milky Way; beyond that, [Fe/H] drops off very 
rapidly with age (see Fig. 13 of Portinari et al. 1998).  The zero point of 
Eq. 1 is consistent with observations of nearby young solar type stars, B 
stars, and H II regions.  The present radial abundance gradient in the thin 
disk, $-0.07$ dex kpc$^{-1}$, is well determined from observations of several 
different kinds of objects in the Milky Way (e.g., B stars, H II regions, 
open clusters, and young solar-type stars -- see Rolleston et al. 2000 
and references cited therein); the uncertainty in the slope given in Eq. 1 is 
based on B star data.  Its magnitude does vary among different galaxy types 
(see Henry and Worthey 1999).  Rolleston et al. (2000) also note that their 
data are consistent with a single slope, as opposed to a sharp break near 10 
kpc from the Galactic center as some have argued previously.  Hou et al. 
(2000) present theoretical estimates of the evolution of radial abundance 
gradients in the Milky Way.  Unfortunately, since observations can only give 
us the present values of the Galactic abundance gradients, estimates of their 
evolution are rather model dependent.  Given this uncertainty and the fact 
that it is a second-order effect, we will ignore possible evolution of the 
gradients in the present study.

In addition to the abundance gradient in the disk, we must also consider the 
spread in [Fe/H] at a given age and R$_{m}$.  The magnitude of the spread 
affects the degree to which the metallicity gradient is an important factor 
for habitability.  If the spread is large in relation to the magnitude of the 
gradient, then the gradient is largely washed out and loses its relevance for 
the radial dependence of habitability in the Milky Way's disk.  A large 
intrinsic spread in the [Fe/H] value of newborn stars at a given R$_{m}$ will 
result in the formation of some stars with a given value of [Fe/H] earlier 
than indicated by Eq. 1.  If, however, the spread is small, then the radial 
metallicity gradient has a more important role in habitability.  The value of 
the spread in [Fe/H] is still a matter of some controversy.  One of the most 
often cited studies is that of Edvardsson et al. (1993), who performed a 
spectroscopic survey of nearly 200 nearby solar-type stars.  They found a 
dispersion in [Fe/H] at a given age of about 0.25 dex, which is much larger 
than observational error alone can explain.  Wielen et al. (1996) modeled the 
observed scatter in [Fe/H] (using the dataset of Edvardsson et al. 1993) 
within the framework of a stellar orbital diffusion model.  Using an improved 
dataset based largely on the Edvardsson et al. (1993) sample, 
Gonzalez (1999a) confirmed the large metallicity spread.

Rocha-Pinto et al. (2000) give an independent estimate of the metallicity 
dispersion in the solar neighborhood with a different sample, finding a value 
about half of the above-cited value.  It is significant that Rocha-Pinto et 
al.'s (2000) estimate of the spread is less than that derived by others.  This 
implies that their sample is more reliable, since the total measured spread is 
the sum of measurement errors, unmodelled biases, and intrinsic cosmic 
scatter; however, it is still possible that an unrecognized systematic bias in 
their analysis may have yielded an artificially small scatter.  Rocha-Pinto et 
al. (2000) attribute the difference largely to selection biases that went into 
the preparation of the original Edvardsson et al. (1993) sample.  Based on 
spectroscopic analyses of young F and G dwarfs, Gonzalez (1999a) and Gaidos and 
Gonzalez (2000) find that the initial dispersion in [Fe/H] is about 0.08 dex; 
the cosmic dispersion must be less than this value, as the measurement error 
is similar in magnitude.  Certainly the last word has not been said on this 
topic; additional observations are required to firmly establish the cosmic 
dispersion.

Since our interest in this section is to estimate the probability of forming 
Earth mass terrestrial planets in the disk of the Milky Way at any time and 
place, the relevant dispersion value in [Fe/H] is that at the time of 
formation of a given star.  For the purpose of the present discussion, then, 
we will adopt a time-constant dispersion in [Fe/H] of 0.08 dex.  We show in 
Figs. 1 and 2 graphical representations of Eq. 1, assuming that terrestrial 
planet mass scales as 10$^{1.5[Fe/H]}$ (see Section 2.1).  The equation has 
not been extrapolated to times within 3 Gyr of the formation of Milky 
Way\footnote{Note, in the present work we assume that the first metals 
were incorporated into the Milky Way were formed 15 Gyr ago.} or 
within 2 kpc of the Galactic center due to the breakdown of Eq. 1 in these 
regimes.  As can be seen in Fig. 2 solar metallicity was reached in the ISM 
in early times at smaller R$_{m}$.

Figs. 1 and 2 should only be considered as rough approximations for reasons 
already given.  Another one is the assumption of constant dispersion in 
metallicity with time.  The cosmic dispersion in metallicity (and in abundance 
ratios) was much greater in the early history of the Milky Way.  This is a 
consequence of the effect on ISM abundances of individual SNe.  During the 
early rapid buildup of metals, the metal-poor ISM was more sensitive to the 
stochastic nature of inhomogeneously distributed SNe.  As the metallicity of 
the ISM increased, individual SNe lead to smaller fractional increases of the 
metallicity (and smaller changes to the abundance ratios).  This effect is 
seen in the abundance patterns of the most metal-poor stars in the halo.  
However, the large observed scatter in abundances among the oldest stars in 
the Milky Way halo is not relevant to the present discussion, as their 
metallicities are too low to form terrestrial planets.  At later times the 
abundance variations can be large for another reason.  As the Milky Way ages, 
the gas mass decreases, and hence, becomes more sensitive to individual SNe 
again.

We argued in Section 2 that certain elements must be present in the right 
proportions to build an Earth-substitute with similar geophysics.  
Specifically, the C/O, Mg/Fe, Si/Fe, and S/Fe ratios are likely to be relevant 
to the habitability of a terrestrial planet.  Of these, high quality 
observational data exist for all but S/Fe.  Below we discuss the observed 
trends in these ratios among thin disk stars.

There is relatively little published observational material on the C/O ratio.  
Garnett et al. (1999) presented C and O abundances for the nearby spiral 
galaxies, M101 and NGC 2403.  They find radial gradients near $-0.05$ dex/kpc, 
though their results are of a preliminary nature, given the small number of 
data points in their sample.  Observations of nearby solar-type stars show a 
significant increase in the C/O ratio with Fe/H ($\sim$0.23 dex/dex) and, 
hence, with time.  The chemical evolution models of Carigi (1996) imply that 
the slope of the C/O ratio with metallicity has steepened greatly in the last 
few Gyrs, finding a value consistent with the observations.  Carigi (1996) 
also finds that the C/O gradient in disk galaxies is negative, consistent with 
the observations of Garnett et al. (1999).

Given these trends, the formation of a planetary system with a C/O ratio 
equivalent to that of the Sun will occur at a specific time for a given 
R$_{m}$.  If a low C/O ratio is required to build habitable planetary systems, 
as was argued in Section 2, then stars born in the solar neighborhood in the 
future might not be accompanied by them.  Without knowing what range of C/O is 
required for habitability, we cannot quantify its effect on the habitable zone 
in the Milky Way, but observations of the parent stars of extra-solar-system 
planets might eventually allow us to better constrain it.

The Mg/Fe and Si/Fe ratios are observed to be declining with increasing Fe/H 
among disk stars with slopes near $-0.3$ and $-0.1$ dex/dex, respectively.  
Therefore, in the future terrestrial planets will form with relatively larger 
iron cores.

\subsubsection{The Thick Disk, Bulge, and Halo}

The thick disk has properties intermediate between the halo and the thin 
disk.  Its constituent stars are only slightly younger that the halo but 
substantially older than the thin disk.  The thick disk population has a 
distinct chemical history compared to the thin disk.  The stars have a mean 
[Fe/H] value near $-0.6$ dex, and most $\alpha$-elements (specifically O, Mg, 
Si, Ca, Ti) have [X/Fe] values 0.2 to 0.3 dex above solar (Gratton et al. 
2000; Prochaska et al. 2000).  The most extreme members reach nearly solar 
[Fe/H], but most $\alpha$-element abundances are relatively constant 
throughout the observed range in [Fe/H].  Hence, if our assumptions from 
Section 2 are correct, then a terrestrial planet that forms around a thick 
disk star is smaller and has a smaller iron core relative to its mantle 
compared to the Earth; the typical terrestrial planet around a thick disk 
star should have about 13\% the mass of the Earth.  However, a small fraction 
of thick disk stars might be accompanied by an Earth-mass terrestrial planet.

The bulge of the Milky Way (or of any other spiral galaxy) is not merely an 
extrapolation of the disk or halo; it is a distinct component.  Star formation 
activity in the bulge peaked earlier than in the disk, but it has continued up 
through the present (Moll· et al. 2000).  Owing to the different chemical 
evolution history of the bulge compared to the disk, the abundance ratios 
among its stars are different from those among stars in the solar 
neighborhood.  Spectroscopic observations by Rich and McWilliam (2000) show 
that [Fe/H] ranges from $-1.6$ to $+0.55$ dex, with the peak near $-0.2$ 
dex.  In addition, they find [X/Fe] values near $+0.2$ to 0.3 dex near solar 
[Fe/H] for the $\alpha$-elements, O, Mg, Si, and Ti. Globular clusters found 
in the bulge appear to share the composition characteristics of the bulge 
stars.  Therefore, Earth-mass terrestrial planets should be common in the 
bulge, but, as with thick disk stars, they likely have relatively small iron 
cores.

Halo stars span a wide range in metallicity, from [Fe/H] = $-4$ to just below 
solar, with the peak near $-1.5$; there is also a trend of decreasing scatter 
in the abundances ratios (expressed as X/Fe) with increasing metallicity (see 
review by Norris 1999).  Star formation was active only during the first 
couple of Gyrs in the halo.  The kinematics result in mostly eccentric, 
plunging orbits, which bring stars close to the Galactic center and pass 
through the disk at high velocity (over 200 km/sec).  The halo also 
contains about 150 globular clusters, which share the same metallicity distribution as the halo field stars, except for the absence of globular 
clusters with [Fe/H] $< -2.5$.  Therefore, on its chemical characteristics 
alone, the halo is very unlikely to contain habitable terrestrial planets.

\subsubsection{Summary of C, O, Mg, Si, Fe}

In summary, each of the four components of the Milky Way is characterized by 
a distinct composition, resulting from a distinct star formation history.  The 
precise composition of the Sun is the result of a history of chemical 
enrichment of the thin disk of the ISM not likely to be found in the other 
components, though cosmic scatter at a given metallicity will lead to some 
overlap.  Even within the thin disk, radial abundance gradients (somewhat 
analogous to the condensation sequence gradient in the early solar nebula) 
ensure that stars with the solar composition are unlikely to form far from 
the solar circle.  Just how large a difference in composition can be tolerated 
in forming an Earth-substitute we cannot yet answer, though.

\subsection{The Long-Lived Radionuclides}
\subsubsection{Evolution of ISM Abundances}

Second in importance for building terrestrial planets, after the mean 
metallicity of the ISM, is likely to be the relative abundances of the 
long-lived radioisotopes.  The heat released from the decay of long-lived 
radioactive isotopes is believed to be the primary source of internal heat in 
the present Earth.  The most important heat-producing isotopes are $^{40}$K, 
$^{232}$Th, $^{238}$U, and $^{235}$U (Fowler 1990; Kargel and Lewis 1993).  
The present abundances of these isotopes have been well determined in 
primitive meteorites (Anders and Grevesse 1989).  Outside the Solar System, 
only the atomic abundances of K and Th can be determined from photospheric 
stellar spectra.  Since $^{232}$Th is the only long-lived isotope of this 
element, its abundance can be equated with the spectroscopically-derived 
atomic value; this cannot be done with K.  We will address the more 
difficult case of $^{40}$K later in this section.

The isotopes of Th and U are produced only by the $r$-process, and SNe II are 
believed to be their main production sites (Mathews et al. 1992; Timmes et al. 
1995; Cowan et al. 1999).  The half-lives of $^{232}$Th, $^{235}$U, and 
$^{238}$U are 14.05, 0.7038, and 4.468 Gyr, respectively.  We begin by 
calculating the evolution of the Th and U abundances relative to Fe in the 
local ISM.  We employ Fe as the reference element for two reasons: 1) Fe is 
the most abundant element in the Earth by weight, and 2) Fe, Th, and U have 
similar condensation temperatures (Lewis 1997).  Therefore, Fe, Th, and U are 
likely to experience fractionation to similar degrees during the early stages 
of planet formation in a protoplanetary disk.

For the purpose of employing theoretical Galactic chemical evolution models, 
it is useful first to compare the evolution of the Th and U abundances to that 
of a stable $r$-process element.  This procedure is advantageous since these 
species are produced by the same source, and there is some empirical evidence 
that the $r$-process abundance pattern is universal (Cowan et al. 1999).  
Assuming it is, the ratio of the abundance of a radioactive $r$-process element 
to that of a stable $r$-process element will have only a weak dependence on the 
details of Galactic chemical evolution.  The best such reference element is 
Eu, because it is easy to measure in stellar spectra, and nearly all of it 
comes from the $r$-process (all but 3\% in the Solar System according to Cowan 
1999).  We can determine the Eu/Fe ratio from observations and convert it to 
Th/Fe and U/Fe ratios by applying theoretical Th/Eu and U/Eu ratios, 
respectively.\footnote{ There are measurements of the Th/Fe ratio in a few 
nearby solar type stars, but they are not considered very reliable, and so 
we will not make use of them.}

We display in Fig. 3 (panel a) the [Eu/Fe] estimates for a sample of 72 nearby 
single F and G dwarfs from the spectroscopic study of Woolf et al. (1995); 
this is currently the best such sample.  We have fit a simple linear equation 
to the data:

\begin{equation}
[Eu/Fe]_{c} = (-0.005\pm0.013) - (0.282\pm0.039)[Fe/H]
\end{equation}

where [Eu/Fe]$_{c}$ is the value of [Eu/Fe] corrected to the solar R$_{m}$ by 
adding the term, 0.11(R$_{m}$ - 8.8), which was determined from a multiple 
linear regression fit to the data with R$_{m}$ and [Fe/H] as independent 
parameters.  A second-order fit would describe the data slightly better, but 
for the present application, a first-order fit is sufficient (see Fig. 3, 
panel b).  When combined with Eq. 1 with R = R$_{m}$, Eq. 2 gives us a 
[Eu/Fe]$_{c}$ - age relation (note that the timescale for this relation is 
based on stellar evolution theory, not nucleocosmochronology).

Next, we calculate [$^{232}$Th, $^{235}$U, $^{238}$U/Eu] - age relations.  In 
this case, observations are very sparse, but theory can help us.  We make use 
of Clayton's (1988) Eqs. (13) and (14), which give the abundances of stable 
and radioactive species in the ISM as a function of time.  The equations are 
based on a Galactic infall model with three adjustable parameters: $k$, 
$\omega$, and $\Delta$.  We reproduce the equations below:

\begin{equation}
Z-Z_{0} = \frac{y \omega \Delta}{k+1} \left[ \frac{t+\Delta}{\Delta} 
- \left( \frac{t+\Delta}{\Delta} \right)^{-k} \right]
\end{equation}

and

\begin{equation}
Z_\lambda-Z_{0}e^{-\lambda t} = y \omega e^{-\lambda t}
\left( \frac{t+\Delta}{\Delta} \right)^{-k} I_{k}(t,\lambda)
\end{equation}

where Eq. 3 applies to stable isotopes and Eq. 4 to radioactive ones.  The 
I$_{k}$(t,$\lambda$) term is given by Clayton as a function of $t$, 
$\Delta$, and $\lambda$ (the radioactive decay rate) for a given $k$.  The 
production yield of a given isotope is denoted by $y$, and the gas consumption 
rate is denoted by $\omega$.  The parameters $\Delta$ and $k$ determine 
infall rate as well as the disk mass growth rate.  We will be working with 
Eqs. 3 and 4 in ratio form only, so knowledge of $\omega$ is not required for 
our application.

Pagel and Tautvaisiene (1995) have applied, in slightly modified form, 
Clayton's model to the observed abundance trends among local disk stars.  They 
derived the following values for the infall model parameters: $k = 3$, 
$\omega = 0.3$ Gyr$^{-1}$, and $\Delta = 4.33$ Gyr; in addition, they adopted 
an age of 15 Gyr for the Milky Way (which is consistent with the timescale of 
our Eq. 1).  These values differ slightly from Clayton's older estimates: 
$k = 1$, $\omega = 0.3$ Gyr$^{-1}$, and $\Delta = 1$ Gyr.  We will adopt the 
estimates of Pagel and Tautvaisiene in the following calculations, since they 
are based on more recent data.  We also need the production ratios.  Cowan 
(1999) gives: $\log (Th/Eu)_{0} = -0.32$, $\log (^{235}U/Eu)_{0} = -0.48$, and 
$\log (^{238}U/Eu)_{0} = -0.68$ (note, combining the theoretical Th/Eu 
production ratio with the observed Th/Eu ratio in two very metal-poor stars, 
Cowan et al. 1999 derive a mean age of $15.6 \pm 4$ Gyr for them).

With this set of parameter values and Eqs. 3 and 4, we have calculated the 
present abundances of the Th and U isotopes relative to Eu for a star formed 
at a time, t, after the formation of the Galaxy (Fig. 4).  The absolute 
abundance ratios have been converted into relative ones using the present 
meteoritic isotopic abundances listed by Anders and Grevesse (1989).  The 
calculations take into account the free decay of the radioactive isotopes in 
the atmosphere of a star between the time of its formation and the 
present.\footnote{Note that Eqs. 3 and 4 yield the abundance of a given 
species in the ISM at some time, t, after the formation of the Milky Way. To 
calculate the free decay of a radioactive isotope between t and the present, 
Eq. 4 must be multiplied by $e^{-(15-t)\lambda}$.} There are a couple of 
points to notice about Fig. 4: 1) the calculations yield very nearly the 
Solar System abundances at t = 10.5 Gyr, and 2) the relative abundances are 
greater in the recently formed stars.  The second point is simply due to the 
fact that the mean age of the radioactive isotopes is less in the younger 
stars.

The close agreement between theory and observation for two very metal-poor 
halo stars and for the Sun gives us confidence that the analysis is 
self-consistent.  But, we should note a word of caution concerning comparison 
of theory with the Solar System abundances.  In general, due to the Weak 
Anthropic Principle, it is dangerous to assume that a particular solar 
parameter is typical (Gonzalez 1999b).  Those studies that use the Solar 
System Th/r-process ratio to derive an estimate of the age of the Milky Way 
might therefore be incorrect.  They should instead use the mean ratio 
determined from a sample of nearby stars which excludes the Sun.  
Unfortunately, it is not possible to determine the Th/Eu ratio in the 
atmosphere of a star as precisely as it can be determined in a meteorite.

To form the [$^{232}$Th, $^{235}$U, $^{238}$U/Fe] - age relations, we can 
combine the results of Fig. 4 with Eqs. (1) and (2).  However, instead of 
using the results of Fig. 4 directly, it is more appropriate for the present 
discussion to calculate the abundances of the radioactive elements 4.5 Gyr 
after the formation of the star.  We show the results of such a calculation 
in Fig. 5.  The results indicate that, relative to Fe, all three radioactive 
isotopes decrease with t.  For example, stars born today will, in 4.5 Gyr, 
have a smaller ratio of Th/Fe than the Sun does today.  This is due to the 
fact that the Fe abundance in the ISM is rising more rapidly than is the Th 
abundance.

The Galactic chemical evolution of $^{40}$K is more difficult to determine for 
two reasons: 1) the sources of $^{40}$K are diverse, both $r$- and $s$-process, 
and 2) it is not possible to measure the abundance of $^{40}$K in the spectra 
of stars.  The most abundant isotopes of K in the Solar System are $^{39}$K and 
$^{41}$K, both of which are produced by Type II SNe (Thielemann et al. 1996); 
stellar spectra only give us the sum of the $^{39}$K, $^{40}$K, and $^{41}$K 
abundances.  In addition, since K is more volatile than U and Th, it will 
condense out of the protoplanetary nebula at relatively larger radial 
distances from the parent star.  Hence, there is a radial gradient in the 
radiogenic isotope concentration among planetary embryos; the gradient should 
be particularly important if significant amounts of K can be incorporated 
into the core, which should form early in the accretion process.

The isotopes of the elements, Se, Kr, Rb, and Sr, are likely to be produced in 
the same site as $^{40}$K (Prantzos et al. 1990; Podosek et al. 1999).  They 
belong to the so-called weak $s$-process component, i.e., the one with the 
smallest neutron exposure, of the Solar System isotopic distribution, which 
are thought to be produced in the He-burning regions of massive stars 
(K\"appeler et al. 1989; Prantzos et al. 1990).  Some $^{40}$K is also 
produced in SNe II (Thielemann et al. 1996), but the primary production is 
probably by the $s$-process.  According to Cowan (1999), the $s$-process 
contribution of K is 89\% in the Solar System.  Thus, although $^{40}$K is 
produced by more than one mechanism, it is probably dispersed into the ISM by 
SNe II (note that we are not saying that $^{40}$K is generated in a SN 
II event, but, rather, that the $^{40}$K produced previously in the interior of 
the SN progenitor is dispersed into the ISM by the explosion).  Of the weak 
$s$-process elements listed above, the easiest to measure in stellar spectra 
is Sr.

Therefore, we will calculate the Galactic evolution of the abundance of 
$^{40}$K relative to that of Sr.  As we did for Th and U, we begin by deriving 
the relation between the stable reference element and Fe.  Employing the 
observations of Gratton and Sneden (1994), we derive:

\begin{equation}
[Sr/Fe] = (-0.18\pm0.04) - (0.29\pm0.06)[Fe/H]
\end{equation}

from 11 stars with [Fe/H] $> -1.3$.  Notably, the slope of this relation is 
nearly identical to that of the [Eu/Fe] trend (Eq. 2), implying that Sr and 
Eu come from the same sources.  However, being a secondary element, one would 
expect Sr to have a different dependence on Fe.  The close agreement between 
the two slopes is probably fortuitous.  As above, we make use of Clayton's 
equations to derive the evolution of the $^{40}$K/Sr ratio.  We adopt the 
value of the half-life for $^{40}$K given by Fowler (1990), 1.25 Gyr, which 
includes the effects of the branched decay of this isotope.  Unfortunately, 
the production ratio of $^{40}$K/Sr ($\equiv$ ($^{40}$K/Sr)$_{0}$) is not well 
known, so, we have set its value to give the present Solar System abundance 
ratio; we get ($^{40}$K/Sr)$_{0} = 0.64$.  The resulting trend of 
[$^{40}$K/Fe] is shown in Fig. 5.  The curve does not pass through zero at 
t $= 10.5$ Gyr, because the Sun has a greater Sr/Fe ratio than other solar 
metallicity stars used to derive Eq. 5; this might be another example of an 
anomalous solar parameter value related to habitability (see Gonzalez 1999b).

Even before we produced Fig. 5, it was already clear that [$^{40}$K/Fe] should 
decline with time given the observed decline of [O/Fe] with time.  This is the 
case because O is injected into the ISM by the same types of stars as is 
$^{40}$K.  However, this does not mean that $^{40}$K is declining in the ISM, 
on the contrary, its mass fraction is increasing (Podosek et al. 1999).

Finally, we should note that our calculations of the time evolution of 
long-lived radioisotopes assume that the evolution of their abundances can 
be approximated by smooth functions.  This assumption is justified if the 
immediately preceding SNe in the vicinity of the forming planetary system 
contributed only a relatively small fraction of freshly synthesized isotopes 
to it.  Podosek et al. (1999) estimate that about 3\% of the $^{40}$K in the 
early Solar Nebula was contributed by a massive star SNe immediately preceding 
its formation.  This is sufficiently small not to invalidate our assumption.

\subsection{Evolution of Terrestrial Planet Radiogenic Heating}

Now that we know the evolution of the abundances of the long-lived 
radionuclides in the ISM, we can estimate the evolution of the radiogenic 
heating in a terrestrial planet.  We will use the Earth as the standard in 
the following discussion for obvious reasons.

In the calculations of the previous section, we adopted what are believed to 
be the isotopic abundances characteristic of the Solar System as a whole 
prior to the fractionation of the volatile elements.  However, the present 
bulk compositions of the planets vary considerably.  In the Earth, the noble 
elements (He, Ne, Ar, etc.) and other volatiles (H, CNO) are greatly 
under-represented relative to the solar photospheric abundances.  Kargel and 
Lewis (1993) give estimates for the amounts of K, Th, and U in the core and 
the mantle+crust of the Earth; they estimate 210 ppm by weight of K and no 
significant U or Th in the core.  More recently, Lodders (1995) and Kargel 
(1995) argue for a higher core abundance of K (550 ppm), based on the 
assumption that the Earth accreted from material similar in composition to 
enstatite chondrites.  We will adopt this second estimate for the core 
abundance of K and Kargel and Lewis's (1993) values for the mantle+crust 
abundances of K, Th, and U.  Herndon (1993) estimate the mass of U in the 
core based on the composition of the Abee enstatite meteorite; he quotes 
masses of 5.8 x 10$^{16}$ g for $^{235}$U and 8.0 x 10$^{18}$ g for $^{238}$U 
in the core, both of which we adopt.  To calculate the mass of $^{232}$Th in 
the core, we assume the same ratio of Th/U as quoted by Anders and Grevesse 
(1999) for meteorites.  This set of abundances results in 
bulk-Earth-to-C1-chondrite ratios of: U $= 1.92$; Th $= 1.89$; K $= 0.62$.  
Relative to Fe, these ratios are: U/Fe $= 1.12$; Th/Fe $= 1.10$; K/Fe $= 
0.36$.

While we adopt specific values for the abundances of the radioactive isotopes 
in this section, we need to be clear that some of these numbers are not yet 
well known.  This is particularly the case for K, for which the core abundance 
is still controversial, and for which the degree of fractionation during 
planet formation is also uncertain.  Therefore, our quoting numbers to two 
significant digits does not imply that the estimates are to be accepted at 
that level of precision.

In order to calculate the radioactive heating, we also need to know the amount 
of energy produced per decay; which we adopt from Faure (1977).  We show the 
results of our calculations for the Earth in Fig. 6.  Our calculations give a 
present-day radiogenic heat production rate of 3.25 x 10$^{20}$ 
erg s$^{-1}$ for the entire Earth; the core contributes 22\%.  The observed 
present global heat loss rate at the Earth's surface is 
($4.42 \pm 0.10$) x 10$^{20}$ erg s$^{-1}$ (Pollack et al. 1993).  It is not 
clear if the deficit can be accounted for entirely from uncertainties in the 
radioisotope abundances.  More likely, other heat sources, such as secular 
cooling of the core and mantle, can make up the difference (see Breuer and 
Spohn 1993 and references cited therein).

Combining the results of this section with those of the previous one, we can 
calculate the evolution of the radiogenic heat production in a terrestrial 
planet formed at any time since the formation of the Milky Way.  There are 
two ways of approaching this problem: 1) we can keep the planet mass constant, 
or 2) we can assume that the planet mass scales with 10$^{1.5[Fe/H]}$ (see 
Section 2.1).  The first case is applicable to a discussion that assumes that 
a terrestrial planet near the Earth's mass is a requirement for complex life.  
The second case is probably closer to reality, but it complicates the 
interpretation, since we are introducing yet another parameter.

As already noted, the bulk composition of the Earth differs from 
C1-chondrites, whose abundance pattern is a close match to that of the solar 
photosphere (except for the most volatile elements).  The differences are 
most likely due to the higher nebular temperatures in the inner 
protoplanetary disk, where the Earth formed.  The place of formation of the 
C1-chondites allowed them to experience relatively less volatile fractionation 
during their condensation.  Therefore, in order to calculate the radiogenic 
heating in the interior of an Earth-like terrestrial planet, we must first 
scale the stellar K/Fe, Th/Fe, and U/Fe abundance ratios from Fig. 5 using 
the bulk Earth-to-C1-chondrite ratios given above.  We show the results of 
such a calculation in Fig. 7.  They indicate that a terrestrial planet 
with the same formation time and age as the Earth produces only 36\% as much 
radiogenic heat, and one forming today will, in 4.5 Gyr, produce only 21\% as 
much.  This is due to the smaller value of $^{40}$K/Fe predicted by our 
Galactic chemical evolution calculations.  It is not clear if the $^{40}$K/Fe 
ratio in the Earth really is anomalously high for its formation time, so we 
have also calculated the radiogenic heating in an Earth-mass terrestrial 
planet assuming the Earth's current radiogenic heat production is typical 
(Fig. 7, dotted curve).  In this second case, an Earth-twin formed today 
will, in 4.5 Gyr, have 60\% of the present Earth's radiogenic heat production.

The radiogenic heat production for variable planet mass is shown in Fig. 
8.  As can be seen in the figure, for a terrestrial planet with a mass that 
scales with 10$^{1.5[Fe/H]}$, the radiogenic heat production levels off.  
However, if we are correct in our speculation that significantly higher 
metallicities lead to less habitable systems (discussed in Section 2), then 
larger terrestrial planets formed in the future at our distance from the 
Galactic center will not necessarily be as habitable as the present Earth.

\section{OTHER GALAXIES}

Like stars, galaxies span a very broad range in luminosity (and mass).  The 
Milky Way is classified as a large spiral galaxy of morphological type 
SAB(rs)bc with an absolute blue magnitude, M$_{\rm B}$, of $-20.2 \pm 0.15$ 
magnitudes (de Vaucouleurs 1982).  It is one of three spirals in the Local 
Group.

It is now well established from observations that a metallicity-luminosity 
correlation exists among galaxies of all types.  Pilyugin and Ferrini (2000) 
discuss this correlation among late-type galaxies (see their Fig. 1); the 
sense of the correlation is such that luminous galaxies are more metal-rich 
than less luminous ones.  We can calculate the fraction of galaxies that are 
less luminous than the Milky Way (and therefore more metal-poor) by comparing 
the Milky Way's luminosity to the nearby galaxy luminosity function.  The 
observed luminosity function of a sample of nearby galaxies is usually fitted 
to a Schechter function (Schechter 1976), which is of the form:

\begin{equation}
\phi (L)dL = \phi^{\star} \left( \frac{L}{L^{\star}} \right)^{\alpha}
exp\left[ -\frac{L}{L^{\star}} \right] \frac{dL}{L^{\star}}
\end{equation}

where

\begin{equation}
\frac{L}{L^{\star}} = 10^{0.4(M^{\star}-M)}
\end{equation}

Folkes et al. (1999) present results of an analysis of a preliminary sample of 
5869 galaxy spectra from the 2dF Galaxy Redshift Survey.  A Schechter function 
fit to their dataset yields the following values for the constants in Eqs. 6 
and 7 (in blue light): M$^{\star} = -19.73$ magnitudes and $\alpha = -1.28$.  
Restricting our calculations to M$_{\rm B}$ between $-14$ and $-22$ 
magnitudes, we find: 1) the Milky Way is among the 0.7\% most luminous 
galaxies in the blue, and 2) these most luminous galaxies contain 16\% of 
the blue light (and, presumably, 16\% of the stars).  Of course, this is only 
an approximate estimate; we did not include in our calculation the cosmic 
dispersion in metallicity at a given luminosity nor the spread in metallicity 
in a given galaxy.  Exclusion of these other factors may increase the 
uncertainty in our estimate by a few percent.

If we were to assume that the total number of stars in a galaxy is 
proportional to its luminosity in the blue, then we could just equate the 
relative luminosities given above to the relative numbers of stars.  
Unfortunately, in the blue we are sensitive to the relatively small number of 
massive blue stars recently formed in a galaxy, so we are in effect biased 
towards galaxies with active star formation activity.  To reduce this bias, 
we can repeat the above calculation with observations obtained with a redder 
bandpass.  Reed (2000) estimates that massive stars contribute about 2\% of 
the Milky Way's luminosity in the visual band.  De Vaucouleurs (1982) gives 
the Milky Way's M$_{\rm V}$ as $-20.73 \pm 0.16$ magnitudes.  Blanton et al. 
(2000) derive luminosity functions in five bandpasses from Sloan Digital Sky 
Survey commissioning data.  None of the Sloan filters is equivalent to the 
Johnson V filter, but the average wavelength of the Sloan g$^{\star}$ and 
r$^{\star}$ filters is equivalent to it.  The Schechter function constants 
corresponding to the average of the Sloan g$^{\star}$ and r$^{\star}$ filters 
are: M$^{\star} = -20.40$ magnitudes and $\alpha = -1.23$.  Restricting our 
calculations to M$_{\rm V}$ between $-15$ and $-23$ magnitudes, we find: 1) 
the Milky Way is among the 1.3\% most luminous galaxies in the visual, and 2) 
these most luminous galaxies contain 23\% of the visual light.

Of course, these results do not apply to the distant (and hence early) 
universe (e.g., the Hubble Deep Fields).  The metallicity was much less then 
(see Wasserburg and Qian 2000 for a discussion of the very early metallicity 
evolution of the universe).  Also, the radiation environment was much more 
hostile due to the higher frequency of transient radiation events (especially 
in the center of a galaxy, where the metallicity is highest).  Therefore, it 
is likely that the Hubble Deep Fields are completely devoid of habitable 
planets (except possibly for a few nearby foreground galaxies).  Until 
recently, studies of the evolution of star formation activity in the larger 
universe indicated that it peaked near a redshift of 2, which corresponds to 
about 3 Gyrs after the Big Bang (Blain and Natarajan 2000).  However, more 
recently, some have begun to argue that the star formation rate peaked prior 
to a redshift of 2 (e.g., Metcalfe et al. 2000).

\section{SUMMARY}

We have argued that there exists a zone of enhanced habitability in the Milky 
Way, which we have termed the Galactic Habitable Zone (GHZ).  One of the more 
important factors is the metallicity of the interstellar matter out of which a 
planetary system forms.  This determines the masses of the terrestrial planets 
in the system (and probably also the gas giants).  This is based on the 
assumption that terrestrial planet mass scales with the surface density of 
solids in a protoplanetary disk, such that lower metallicity leads to smaller 
planets.  We estimate, very approximately, that a metallicity at least half 
that of the Sun is required to build a habitable terrestrial planet.  These 
assumptions need to be tested with simulations of planetary formation with 
adjustable initial metallicity starting with the collapse of the birth 
cloud.

Also important is the decrease in the interstellar medium abundances of the 
long-lived radioisotopes, $^{40}$K, $^{235,238}$U, and $^{232}$Th relative to 
Fe.  Given the lack of a suitable alternative geothermal energy source, 
radiogenic heating is a necessary requirement for the long-term maintenance of 
a terrestrial planet's habitability via climate stability provided by the 
carbon cycle.  Of somewhat lesser importance for habitability are the ratios: 
C/O, Si/Fe, Mg/Fe, and S/Fe.  These affect the water content, the core to 
mantle mass ratio, and the state of the core of a terrestrial planet.  The 
least certain aspect of our calculations concerns the abundance of $^{40}$K, 
both in the interstellar medium and in terrestrial planets.  Additional 
research on all aspects of $^{40}$K abundances is warranted.

In the process of defining the GHZ, we found it necessary first to establish 
the links between the Circumstellar Habitable Zone (CHZ) and Galactic scale 
phenomena.  As a result of our review of studies of the CHZ, we found that 
many important factors are not included in its general definition.  These 
include the radial composition gradient in the early protoplanetary disk 
(especially with regard to $^{40}$K), the radial dependence of impactor 
energy, and details of the late stages of planet accretion.

Based on Galactic chemical evolution alone, we find, not surprisingly, that 
the thin disk near the Sun is the most likely place for Earth-like planets to 
form in the present time.  On average, the inner disk should contain 
terrestrial planets larger than the Earth, and the outer disk is likely to 
contain smaller terrestrial planets.  Given recent evidence of a metallicity 
dependence on giant planet formation, they should also be more common in the 
inner Galactic disk.  The bulge should contain many Earth-mass planets but 
relatively few Earth-like planets, given the different mix of elements among 
its stars.  The evolving concentration of the geophysically important 
radioisotopes in the ISM establishes a window of time (albeit with fuzzy 
boundaries) in the history of the Milky Way during which terrestrial 
habitable planets with long-lasting geological activity can exist.  That 
window is slowly closing with a timescale of billions of years.

Observations of other galaxies show that the Milky Way's star formation 
history is probably not atypical.  Star formation in other large galaxies 
probably peaked at about the same time as in the Milky Way.  The observed 
correlation in the nearby universe between the metallicity of a galaxy and 
its luminosity implies that low luminosity galaxies are unlikely to contain 
Earth-mass planets yet.

Of course, the definition of the GHZ presented in this study is not complete, 
as we did not discuss Galactic scale constraints not directly related to 
Galactic chemical evolution.  These include threats from transient radiation 
events and dynamical perturbations of comet and planet orbits (to be discussed 
in future papers in this series).  Consideration of these other constraints 
lead us to exclude the bulge and inner disk from the GHZ.  To summarize, then, 
the GHZ is an annulus in the thin disk of the Milky Way that migrates outward 
with time as the heavy elements build up.

\acknowledgments

The authors thank J. J. Cowan, J. Lissauer, and G. Wallerstein for helpful 
advice and comments.  The constructive and thoughtful critiques of the 
reviewers, J. Lissauer, C. P. McKay, and V. Trimble, helped to clarify the 
presentation and are greatly appreciated.  GG acknowledges support from the 
Kennilworth Fund of the New York Community Trust.

\clearpage

\figcaption{Terrestrial planet mass is plotted against time since the 
formation of the Milky Way for the ISM in the solar neighborhood (diagram 
{\bf a}).  The nominal trend calculated from Eq. 1 is shown as a solid curve; 
mass of planet is assumed proportional to 10$^{1.5[Fe/H]}$ (see Section 2.1).  
The one sigma upper and lower bounds are shown as dotted curves.  The Sun is 
shown as an open circle.  Terrestrial planet mass is plotted against 
Galactocentric distance for the present ISM (diagram {\bf b}).}

\figcaption{Locus of points on the time-Galactocentric distance plane 
corresponding to solar metallicity of the ISM.  The diagram is divided into 
metal-poor and metal-rich regions (with respect to solar metallicity). Symbol 
meanings are the same as in Fig. 1.}

\figcaption{Trend of [Eu/Fe] with [Fe/H] using data from Woolf et al. (1995), 
corrected for R$_{\rm m} = 8.8$ kpc (dots; panel {\bf a}); the bracket 
notation employed here and in the following captions is a logarithmic 
abundance scale relative to the Sun (see text for a definition).  The data 
have been separated into three groups in panel {\bf b}; the standard error of 
the mean is shown for each point.  Equation 2 is shown as a line in panel 
{\bf b}.}

\figcaption{The present relative abundances of the $r$-process radioisotopes, 
as [$^{232}$Th, $^{235}$U, $^{238}$U/Eu], as a function of a star's formation 
time, t.  The abundances shown are relative to the present meteoritic 
abundances from Anders and Grevesse (1989). The Solar System value is 
indicated with an open circle.}

\figcaption{The [$^{232}$Th, $^{235}$U, $^{238}$U, $^{40}$K/Fe] values in a 
star 4.5 Gyr after its formation as a function of formation time, t.  The 
error bars shown at the time of the formation of the solar system are based 
on an uncertainty of $\pm$ 10\% in the production ratios.  The solar system 
value is indicated with an open circle.}

\figcaption{Evolution of radiogenic heating from the decay of 
$^{40}$K, $^{232}$Th, $^{235}$U, and $^{238}$U in the Earth's interior.  
The dot represents the observed present heat loss at the surface of the Earth 
(Pollack et al. 1993). Here t corresponds to the time since the formation of 
the Earth.}

\figcaption{Radiogenic heating in an Earth-mass planet 4.5 Gyr after its 
formation relative to the present radiogenic heating in the Earth as a 
function of formation time (solid curve).  This is based on the data in 
Figs. 5 and 6.  The data have also been multiplied by a constant in order to 
pass through unity at t = 10.5 Gyr (dotted curve).}

\figcaption{Same as Fig. 7 but for planet mass that is proportional to 
10$^{1.5[Fe/H]}$; the solid and dotted curves have the same meaning as 
before.  The planet mass relative to the Earth is indicated by a dashed curve.}


\clearpage

\begin{thebibliography}{}
\bibitem[]{}Anders, E., and N. Grevesse 1989. Abundances of the elements: 
meteoritic and solar. Geochim. et Cosmochem. Acta {\bf 53}, 197-214.
\bibitem[]{}Balazs, B. S. 1988. The Galactic belt of intelligent life. In 
Bioastronomy -- The next steps (G. Max, Ed.), pp. 61-66. Kluwer Academic Pub., 
Dordrecht.
\bibitem[]{}Blain, A. W., and P. Natarajan 2000. Gamma-ray bursts and the 
history of star formation. \mnras~{\bf 312}, L35-38.
\bibitem[]{}Blanton, M. R., et al. 2001. The luminosity function of galaxies 
in SDSS commissioning data. \aj, submitted.
\bibitem[]{}Brakenridge, G. R. 1981. Terrestrial paleoenvironmental effects of 
a late quaternary-age supernova. Icarus~{\bf 46}, 81-93.
\bibitem[]{}Breuer, D., and T. Spohn 1993. Cooling of the Earth, Urey ratios, 
and the problem of potassium in the core. Geophys. Res. Let. {\bf 20}, 
1655-1658.
\bibitem[]{}Buffett, B. A. 2000. Earth's core and the geodynamo. Science {\bf 
288}, 2007-2012.
\bibitem[]{}Butler, R. P. et al. 2000. Statistical properties of extrasolar 
planets. In Planetary systems in the universe: Observation, formation, and 
evolution (Penny A, Artymowicz P, Lagrange A -M, and Russell, Eds.), ASP 
Conference Series, in press.
\bibitem[]{}Butler, R. P. et al. 2001. Two new planets from the 
Anglo-Australian planet search. \aj, submitted.
\bibitem[]{}Carigi, L. 1996. Models of Chemical Evolution of the Galactic 
Disk: Considering Different SFR, IMF, and Infall Rates. Rev. Mexicana Astron. 
Astrofis. {\bf 32}, 179-192.
\bibitem[]{}Chen, Y. Q., P. E. Nissen, G. Zhao, H. W. Zhang, and T. Benoni 
2000. Chemical composition of 90 F and G dwarfs. \aaps~{\bf 141}, 491-506.
\bibitem[]{}Clarke, J. N. 1981. Extraterrestrial intelligence and galactic 
nuclear activity. Icarus {\bf 46}, 94-96.
\bibitem[]{}Clayton, D. D. 1988. Nuclear cosmochronology within analytic 
models of the chemical evolution of the solar neighborhood. \mnras~{\bf 234}, 
1-36.
\bibitem[]{}Cowan, J. J. 1999, private communication.
\bibitem[]{}Cowan, J. J., A. McWilliam, C. Sneden, and D. L. Burris 1997. The 
thorium chronometer in CS 22892-052: estimates of the age of the Galaxy. 
\apj~{\bf 480}, 246-254.
\bibitem[]{}Cowan, J. J., B. Pfeiffer, K.-L. Kratz, F.-K. Thielemann, C. 
Sneden, S. Burles, D. Tytler, and  T. Beers 1999.  R-process abundances and 
chronometers in metal-poor stars. \apj~{\bf 521}, 194-205.
\bibitem[]{}de Vaucouleurs, G. 1982. Five crucial tests of the cosmic distance 
scale using the Galaxy as fundamental standard. Proc. Astron. Soc. Aus. {\bf 
4(4)}, 320-327.
\bibitem[]{}Draganic, I. G., E. Bjergbakke, Z. D. Draganic, and K. Sehested 
1991. Decomposition of ocean waters by potassium-40 radiation 3800 Ma ago as 
a source of oxygen and oxidizing species. Precam. Res. {\bf 52}, 337-345.
\bibitem[]{}Edvardsson, B., J. Andersen, B. Gustafsson, D. L. Lambert, P. E. 
Nissen, and J. Tomkin 1993. The chemical evolution of the Galactic disk: I. 
analysis and results. \aap~{\bf 275}, 101-152.
\bibitem[]{}Ellis, J. and D. N. Schramm 1995. Could a nearby supernova 
explosion have caused a mass extinction? Proc. Nat. Acad. Sci. {\bf 92}, 
235-238.
\bibitem[]{}Faure, G. 1977. Principles of Isotope Geology, John Wiley \& Sons, 
New York.
\bibitem[]{}Folkes, S. et al. 1999. The 2dF Galaxy redshift survey: spectral 
types and luminosity functions. \mnras~{\bf 308}, 459-472.
\bibitem[]{}Fowler, C. M. R. 1990. The Solid Earth: An Introduction to Global 
Geophysics, Cambridge Univ. Press, Cambridge.
\bibitem[]{}Franck, S. 1998. Evolution of the global mean heat flow over 4.6 
Gyr. Tectonophysics {\bf 291}, 9-18.
\bibitem[]{}Franck, S., A. Block, W. von Bloh, C. Bounama, H. J. Schellnhuber, 
and Y. Svirezhev 2000a. Reduction of biosphere life span as a consequence of 
geodynamics. Tellus {\bf 52B}, 94-107.
\bibitem[]{}Franck, S., A. Block, W. von Bloh, C. Bounama, M. Steffen, D. 
Schonberner, and H. J. Schellnhuber 2000b. Determination of habitable zones 
in extrasolar planetary systems: where are Gaia's sisters? JGR {\bf 105}, 
1651-1658.
\bibitem[]{}Franck, S., K. Kossacki, and C. Bounama 1999. Modelling the global 
carbon cycle for the past and future evolution of the Earth system. Chem. 
Geol. {\bf 159}, 305-317.
\bibitem[]{}Gaidos, E. J. 2000. A cosmochemical determinism in the formation 
of Earth-like planets. Icarus {\bf 145}, 637-640.
\bibitem[]{}Gaidos, E. J., and G. Gonzalez 2000. Young solar analogs. in 
preparation.
\bibitem[]{}Garnett, D. R., et al. 1999. Carbon in Spiral Galaxies from Hubble 
Space Telescope Spectroscopy. \apj~{\bf 513}, 168-179.
\bibitem[]{}Gilliland, R. L. et al. 2000. A lack of planets in 47 Tucanae from 
an HST search. \apj~{\bf 545}, L47-51.
\bibitem[]{}Gonzalez, G. 1999a. Are stars with planets anomalous? \mnras~{\bf 
308}, 447-458.
\bibitem[]{}Gonzalez, G. 1999b. Is the sun anomalous? Astron. \& Geophys. {\bf 
40}, 5.25-5.29.
\bibitem[]{}Gonzalez, G., C. Laws, S. Tyagi, and B. E. Reddy 2001. Parent 
stars of extrasolar planets VI: Abundance analyses of 20 new systems. \aj~{\bf 
121}, 432-452.
\bibitem[]{}Goswami, A., and N. Prantzos 2000. Abundance evolution of 
intermediate mass elements (C to Zn) in the Milky Way halo and disk. \aap~{\bf 
359}, 191-212.
\bibitem[]{}Gratton, R. G., and C. Sneden 1994. Abundances of neutron-capture 
elements in metal-poor stars. \aap~{\bf 287}, 927-946.
\bibitem[]{}Gratton, R. G., E. Carretta, F. Matteucci, and C. Sneden 2000. 
Abundances of light elements in metal-poor stars. IV. [Fe/O] and [Fe/Mg] 
ratios and the history of star formation in the solar neighborhood. \aap~{\bf 
358}, 671-681.
\bibitem[]{}Greenberg, R., P. Geissler, B. R. Tufts, and G. Hoppa 2000. 
Habitability of Europa's crust: the role of tidal-tectonic processes. J. 
Geophys. Res. {\bf 105}, 17,551-17,562.
\bibitem[]{}Hart, M. H. 1979. Habitable zones about main sequence stars. 
Icarus {\bf 37}, 351-357.
\bibitem[]{}Henry, R. B. C., and G. Worthey 1999. The distribution heavy 
elements in spiral and elliptical galaxies. \pasp~{\bf 111}, 919-945.
\bibitem[]{}Henry, T. J. 1998. Suspicious Characters Lurking in the Solar 
Neighborhood. In Brown Dwarfs and Extrasolar Planets (Rebolo R, Martin E L, 
Osorio M R Z, Eds.), pp. 28-35. Book Crafters, San Francisco.
\bibitem[]{}Henry, T. J. 1999. private communication.
\bibitem[]{}Herndon, J. M. 1993. Feasibility of a nuclear fission reactor at 
the center of the Earth as the energy source of the geomagnetic field. J. 
Geomag. Geoelectric. {\bf 45}, 423-437.
\bibitem[]{}Hou, J. L, N. Prantzos, and S. Boissier 2000. Abundance gradients 
and their evolution in the Milky Way disk. \aap~{\bf 362}, 921-936.
\bibitem[]{}Huang, S.-S. 1959. Occurrence of life in the universe. American 
Scientist {\bf 47}, 397-402.
\bibitem[]{}Kappeler, F., H. Beer, and K. Wisshak 1989. $s$-process nucleosynthesis - nuclear physics and the classical model. Rep. Prog. Phys. 
{\bf 52}, 945-1013.
\bibitem[]{}Karato, S.-I., and H. Jung 1998. Water, partial melting and the 
origin of the seismic low velocity and high attenuation zone in the upper 
mantle. Earth Plan. Sci. Lett. {\bf 157}, 193-207.
\bibitem[]{}Kargel, J. S. 1995. A possible enstatite meteorite-earth 
connection and potassium in Earth's core. Meteoritics {\bf 30}, 5-7.
\bibitem[]{}Kargel, J. S., and J. S. Lewis 1993. The composition and early 
evolution of Earth. Icarus {\bf 105}, 1-25.
\bibitem[]{}Kasting, J. F., D.P. Whitmire, R.T. Reynolds 1993. Habitable 
zones around main sequence stars. Icarus {\bf 101}, 108-128.
\bibitem[]{}Lewis, J. R., and K. C. Freeman 1989. Kinematics and chemical 
properties of the old disk of the galaxy. \aj~{\bf 97}, 139-162.
\bibitem[]{}Lewis, J. S. 1997. Physics and Chemistry of the Solar System. 
Academic Press, San Diego, CA.
\bibitem[]{}Lewis, J. S. 1998. Worlds Without End: the exploration of planets 
known and unknown. Perseus Books, Reading, MA.
\bibitem[]{}Lineweaver, C. H. 2000. An estimate of the age distribution of terrestrial planets in the universe: quantifying metallicity as a selection 
effect. Icarus, in press.
\bibitem[]{}Lissauer, J. J. 1995. Urey prize lecture: on the diversity of 
plausible planetary systems. Icarus {\bf 114}, 217-236.
\bibitem[]{}Lodders, K. 1995. Alkali elements in the Earth's core: evidence 
from enstatite meteorites. Meteoritics {\bf 30}, 93-101.
\bibitem[]{}Lodders, K. and B. Fegley, Jr. 1998. The Planetary Scientist's 
Companion. Oxford University Press, New York.
\bibitem[]{}Marochnik, L. S. 1984. On the position of the Sun in the Galaxy. 
Astrophysics {\bf 19}, 278-283.
\bibitem[]{}Mathews, G. J., G. Bazan, and J. J. Cowan 1992. Evolution of 
heavy-element abundances as a constraint on sites for neutron-capture 
nucleosynthesis. \apj~{\bf 391}, 719-735.
\bibitem[]{}McKay, C. P. 1996. Time for intelligence on other planets. In 
Circumstellar Habitable Zones: proceedings of the first international 
conference (Doyle L R, Ed.), pp. 405-419. Travis House Publications, Menlo 
Park, CA.
\bibitem[]{}McWilliam, A. 1997. Abundance ratios and Galactic chemical 
evolution. \araa~{\bf 35}, 503-556.
\bibitem[]{}Metcalfe, N., T. Shanks, A. Campos, H. J. McCracken, and R. Fong 
2000. Galaxy number counts -- V. Ultra-deep counts: The Herschel and Hubble 
Deep Fields. \mnras, submitted.
\bibitem[]{}Molla, M., F. Ferrini, and G. Gozzi 2000. Galactic bulges. 
\mnras~{\bf 316}, 345-356.
\bibitem[]{}Norris, J. E. 1999. The Chemical Abundance Structure of the 
Galactic Halo. In The third Stromlo Symposium: The Galactic Halo (Gibson B K, 
Axelrod T S, and Putnam M E, Eds.), pp. 213-224. Book Crafters, San Francisco.
\bibitem[]{}Moulton, K. L., and R. A. Berner 1998. Quantification of the 
effect of plants on weathering; studies in Iceland. Geology {\bf 26(10)}, 
895-898.
\bibitem[]{}Pagel, B. 1997. Nucleosynthesis and Galactic Chemical Evolution. 
Cambridge University Press, Cambridge.
\bibitem[]{}Pagel, B. E. J., and G. Tautvaisiene 1995. Chemical evolution of 
primary elements in the Galactic disk: an analytical model. 
\mnras~{\bf 276}, 505-514.
\bibitem[]{}Pilyugin, L. S., and F. Ferrini 2000. On the origin of the 
luminosity-metallicity relation for late-type galaxies: spirals to irregulars 
transition. \aap~{\bf 358}, 72-76.
\bibitem[]{}Podosek, F. A., R. H. Nichols, Jr., J. C. Brandon, B. S. Meyer, U. 
Ott, C. L. Jennings, and N. Luo 1999. Potassium, stardust, and the last 
supernova. Geochimica et Cosmochemica Acta {\bf 63}, 2351-2362.
\bibitem[]{}Pollack, H. N., S. J. Hurter, J. R. Johnson 1993. Heat flow from 
the Earth's interior -- analysis of the global data set. Rev. Geophys. {\bf 
31(3)}, 267-280.
\bibitem[]{}Portinari, L., C. Chiosi, and A. Bressan 1998. Galactic chemical 
enrichment with new metallicity dependent stellar yields. \aap~{\bf 334}, 
505-539.
\bibitem[]{}Pratzos, N., M. Hashimoto, and K. Nomoto 1990. The s-process in 
massive stars: yields as a function of stellar mass and metallicity. \aap~{\bf 
234}, 211-229.
\bibitem[]{}Prochaska, J. X., S. O. Naumov, B. W. Carney, A. McWilliam, and A. 
M. Wolfe 2000. The Galactic thick disk stellar abundances. \aj~{\bf 120}, 
2513-2549.
\bibitem[]{}Reed, B. C. 2000. New estimates of the scale height and surface 
density of OB stars in the solar neighborhood. \aj~{\bf 120}, 314-318.
\bibitem[]{}Rich, R. M., and A. McWilliam 2000. Abundances of stars in the 
galactic bulge obtained using the Keck Telescope. Proc. SPIE {\bf 4005}, 
150-161.
\bibitem[]{}Rocha-Pinto, H. J. and W. J. Maciel 1996. The metallicity 
distribution of G dwarfs in the solar neighborhood. \mnras~{\bf 279}, 447-458.
\bibitem[]{}Rocha-Pinto, H. J., W. J. Maciel, J. Scalo, and C. Flynn 2000. 
Chemical enrichment and star formation in the Milky Way. I. Sample description 
and chromospheric age-metallicity relation. \aap~{\bf 358}, 850-868.
\bibitem[]{}Rolleston, W. R. J., S. J. Smartt, P. L. Dufton, and R. S. I. 
Ryans 2000. The Galactic metallicity gradient. \aap~{\bf 363}, 537-554.
\bibitem[]{}Sagan, C., and J. S. Shklovsky 1966. Intelligent life in the 
universe. Holden-Day, San Francisco.
\bibitem[]{}Samland, M. 1998. Modeling the evolution of disk galaxies. II. 
Yields of massive stars. \apj~{\bf 496}, 155-171.
\bibitem[]{}Schechter, P. L. 1976. An analytic expression for the luminosity 
function of galaxies. \apj~{\bf 203}, 297-306.
\bibitem[]{}Schenk, P. M., E. Asphaug, W. B. McKinnon, H. J. Melosh, and P. R. 
Weissman 1996. Cometary nuclei and tidal disruption: the geologic record of 
crater chains on Callisto and Ganymede. Icarus {\bf 121}, 249-274.
\bibitem[]{}Tackley, P. J. 2000. Mantle convection and plate tectonics: toward 
an integrated physical and chemical theory. Science {\bf 288}, 2002-2007.
\bibitem[]{}Thielemann, F.-K., K. Nomoto, and M. Hashimoto 1996. Core-collapse 
supernovae and their ejecta. \apj~{\bf 460}, 408-436.
\bibitem[]{}Timmes, F. X., S. E. Woosley, and T. A. Weaver 1995. Galactic 
chemical evolution: Hydrogen through Zinc. \apjs~{\bf 98}, 617-658.
\bibitem[]{}Trimble, V. 1997a. Origin of the biologically important 
elements.Origins of Life Evol. Biosphere {\bf 27}, 3-21.
\bibitem[]{}Trimble, V. 1997b. Galactic chemical evolution: implications for 
the existence of habitable planets. In Extraterrestrials -- Where are They?  
(Zuckerman B. \& Hart M., Eds.), pp. 184-191. Cambridge University Press, 
Cambridge.
\bibitem[]{}Tucker, W. C. 1981. Astrophysical crises in the evolution of life 
in the Galaxy. In Life in the Universe (Billingham J., Ed.), pp. 287-296. The 
MIT Press, Cambridge.
\bibitem[]{}Ward, P., and D. Brownlee 2000. Rare Earth. Copernicus, New York.
\bibitem[]{}Wasserburg, G. J., and Y.-Z. Qian 2000. A model of metallicity 
evolution in the early universe. \apj~{\bf 538}, L99-L102.
\bibitem[]{}Weidenschilling, S. J., and F. Marzari 1996. Gravitational 
scattering as a possible origin for giant planets at small distances. Nature 
{\bf 384}, 619-621.
\bibitem[]{}Wetherill, G. W. 1996. The formation and habitability of 
extra-solar planets. Icarus {\bf 119}, 219-238.
\bibitem[]{}Whitmire, D. P. and R. T. Reynolds 1996. Circumstellar habitable 
zones: astronomical considerations. In Circumstellar Habitable Zones (L. R. 
Doyle, Ed.), pp. 117-142. Menlo Park, Travis House Publications.
\bibitem[]{}Whittet, D. 1997. Astronomy \& Geophys. {\bf 38(5)}, 8.
\bibitem[]{}Williams, D. M., J. F. Kasting, and R. A. Wade 1997. Habitable 
moons around extrasolar giant planets. Nature {\bf 385}, 234-236.
\bibitem[]{}Woolf, V. M., J. Tomkin, and D. L. Lambert 1995. The r-process 
element Europium in Galactic disk F and G dwarf stars. \apj~{\bf 453}, 660-672.
\end{thebibliography}
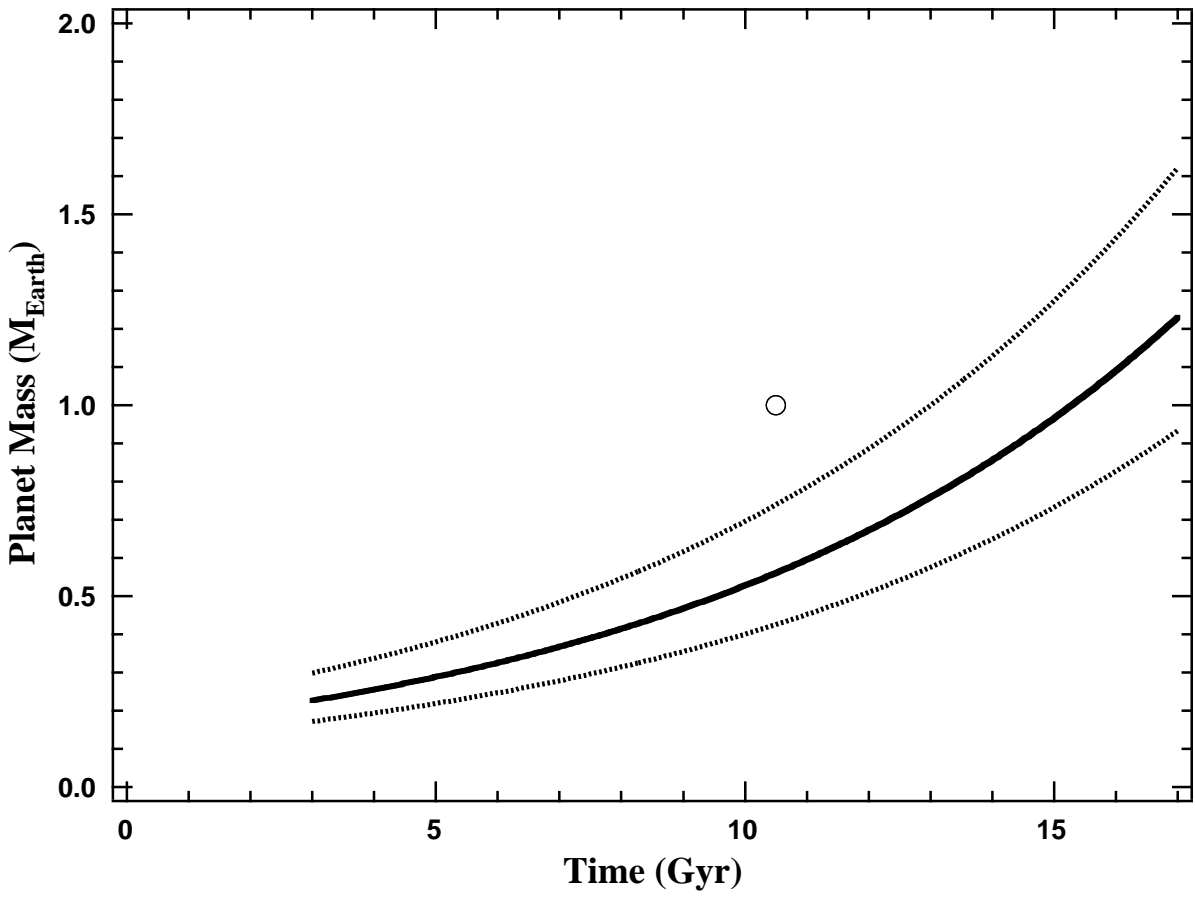
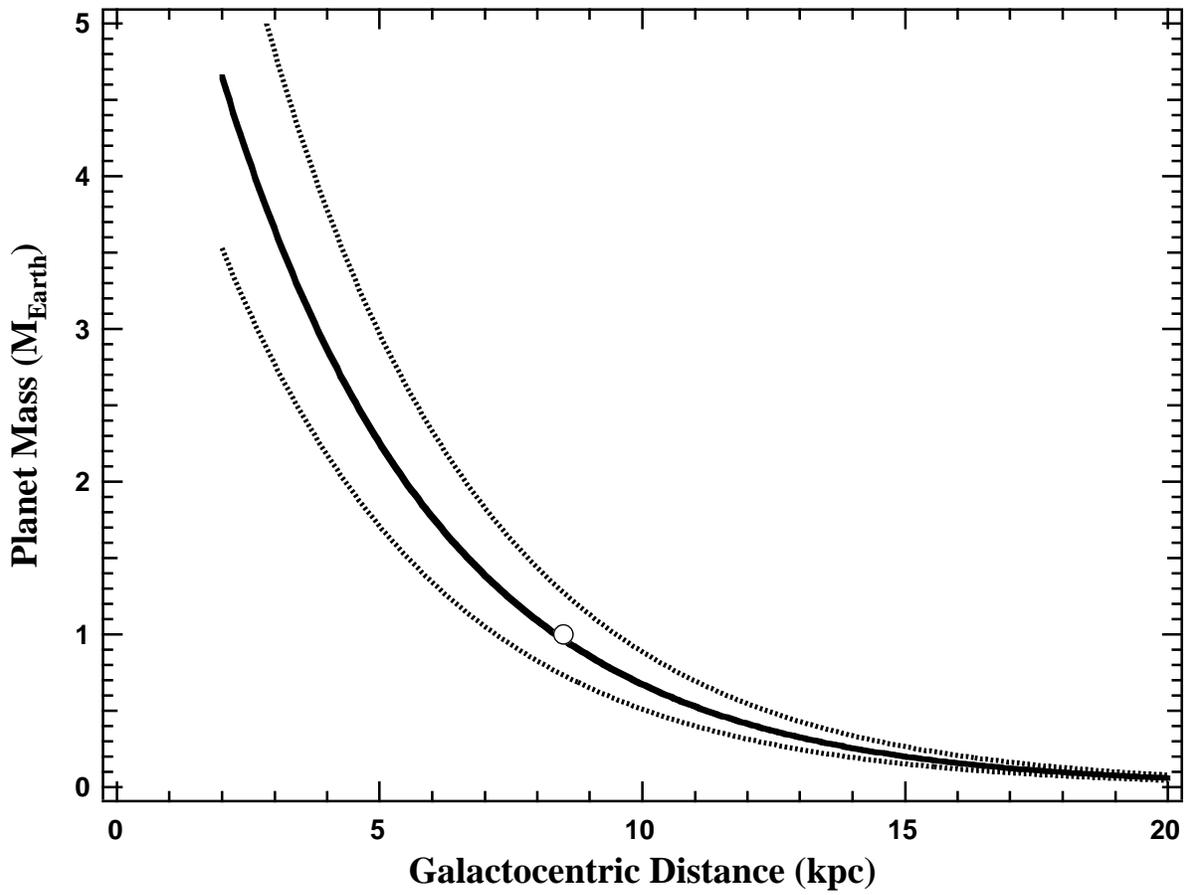
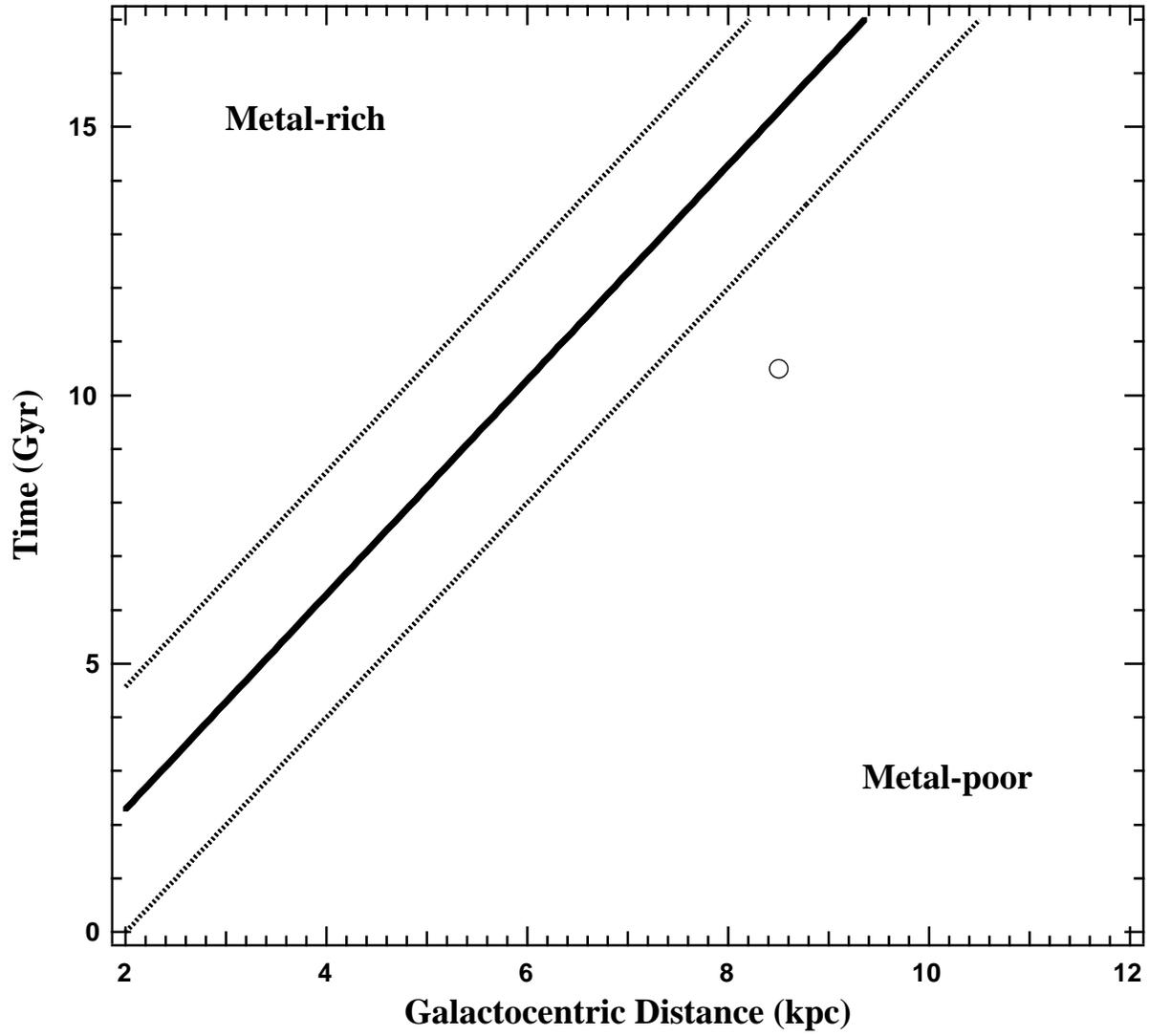
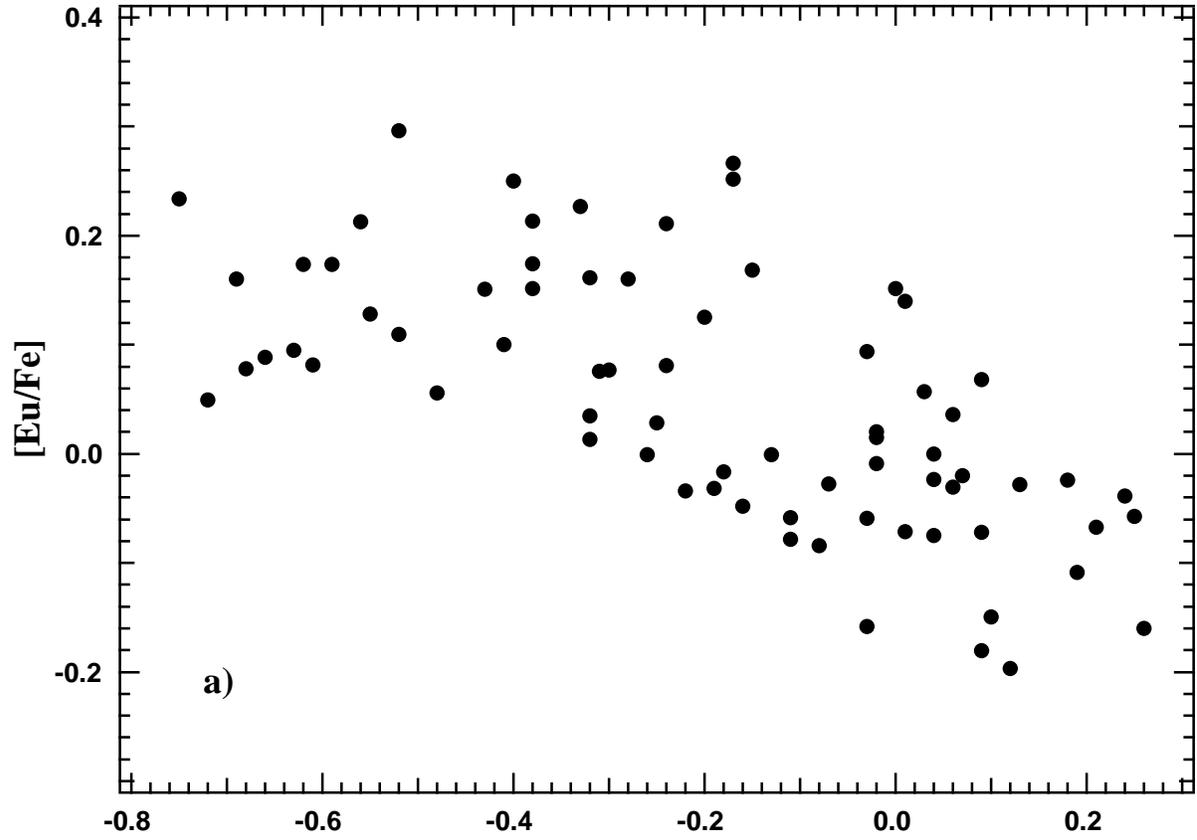
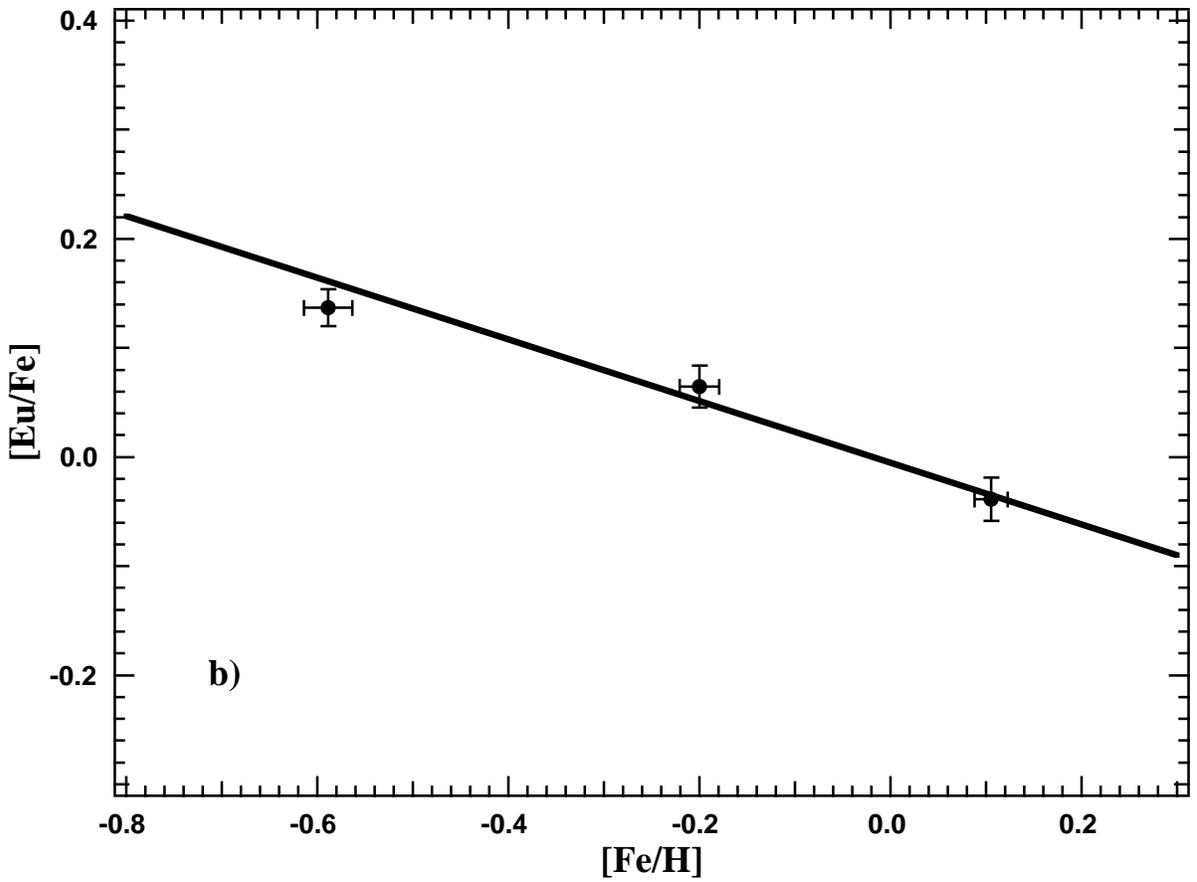
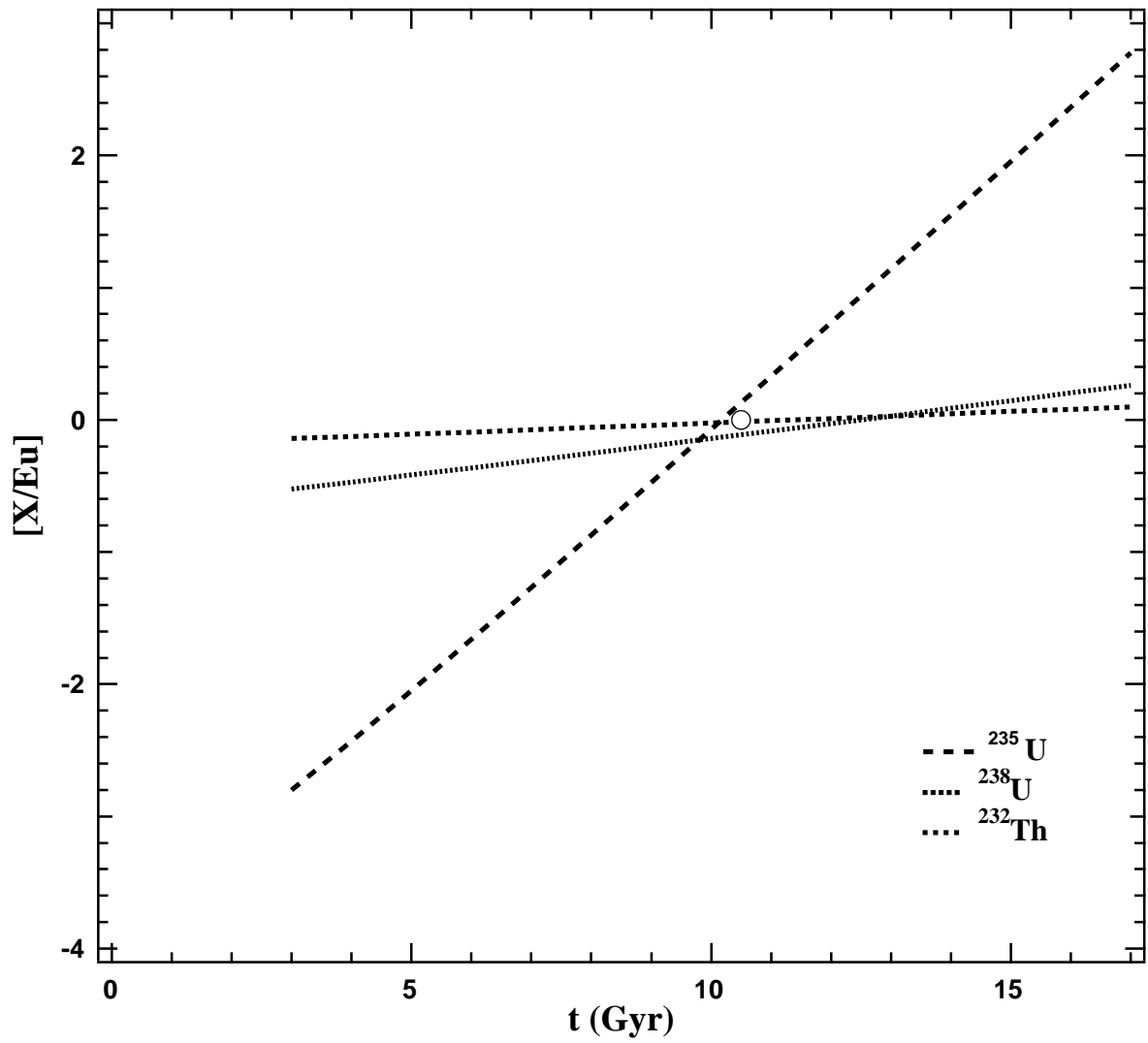
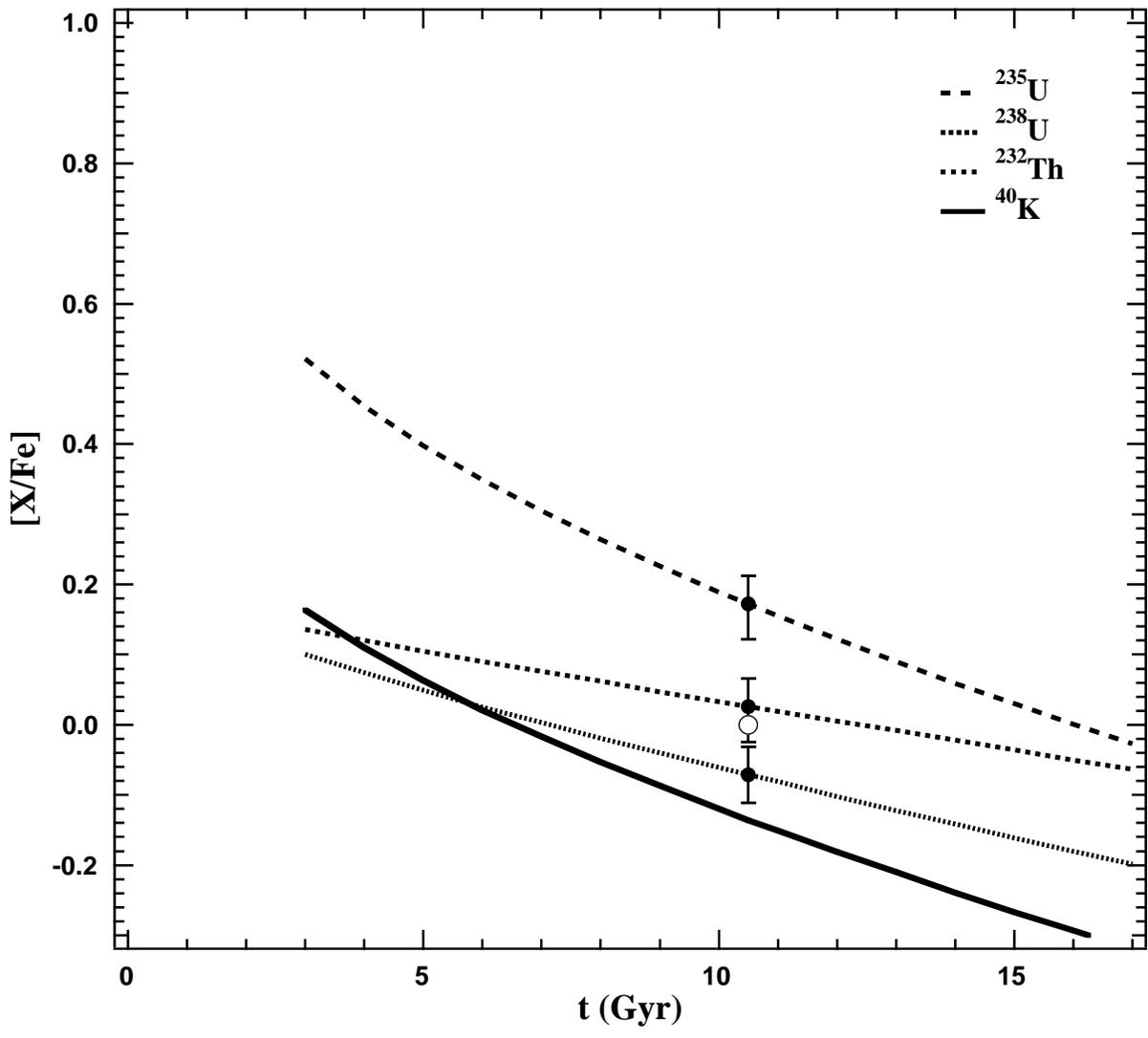
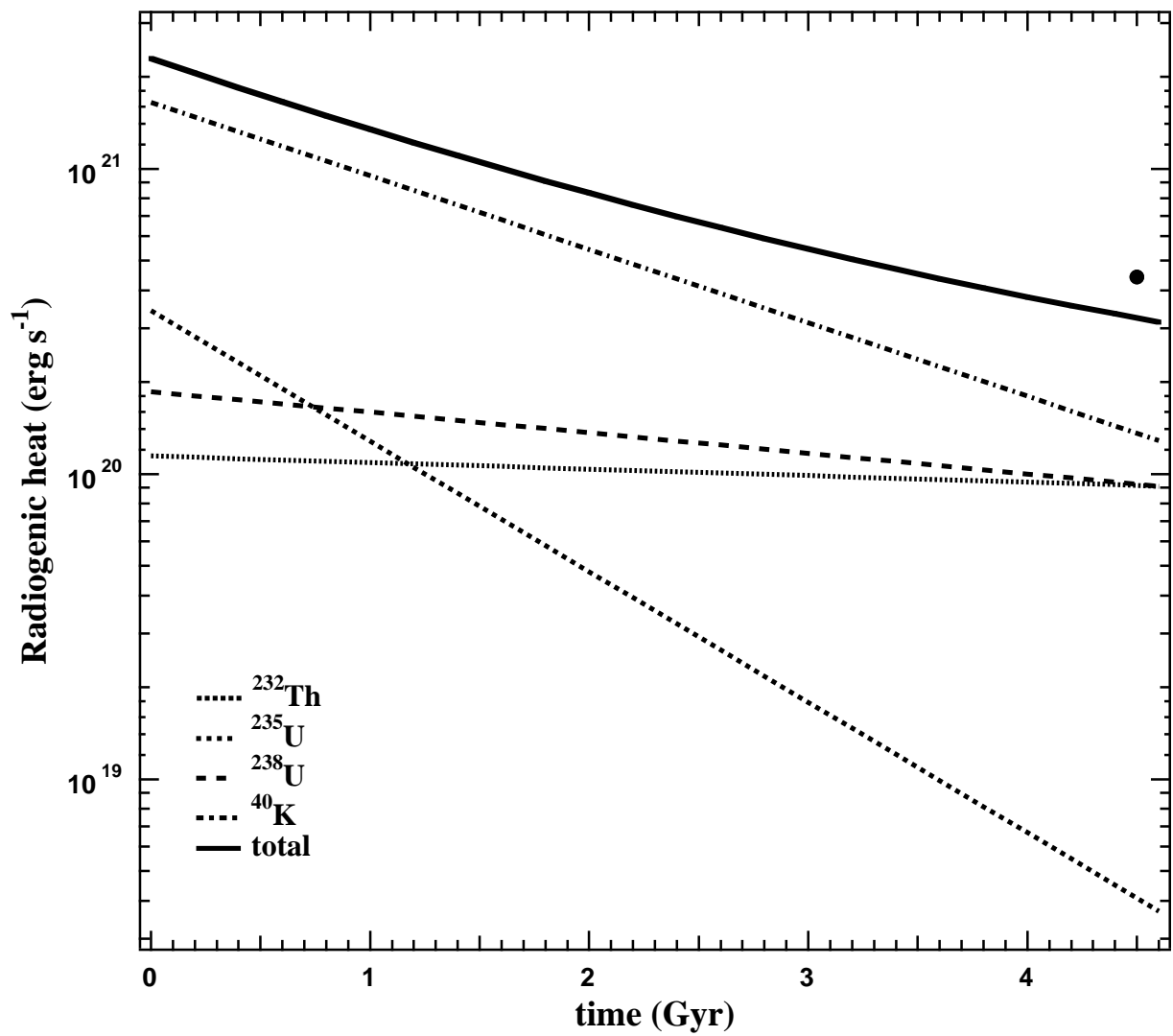
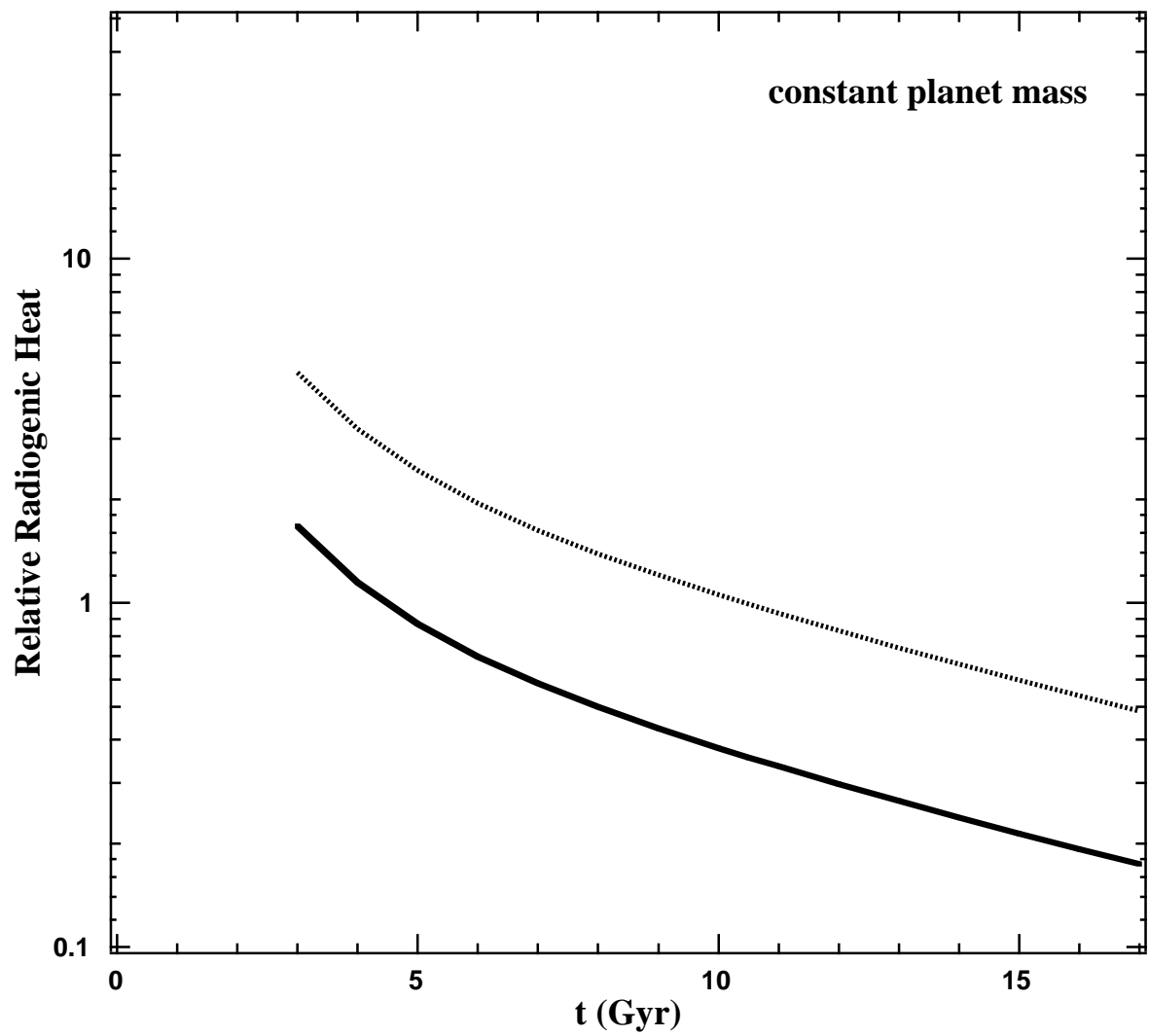
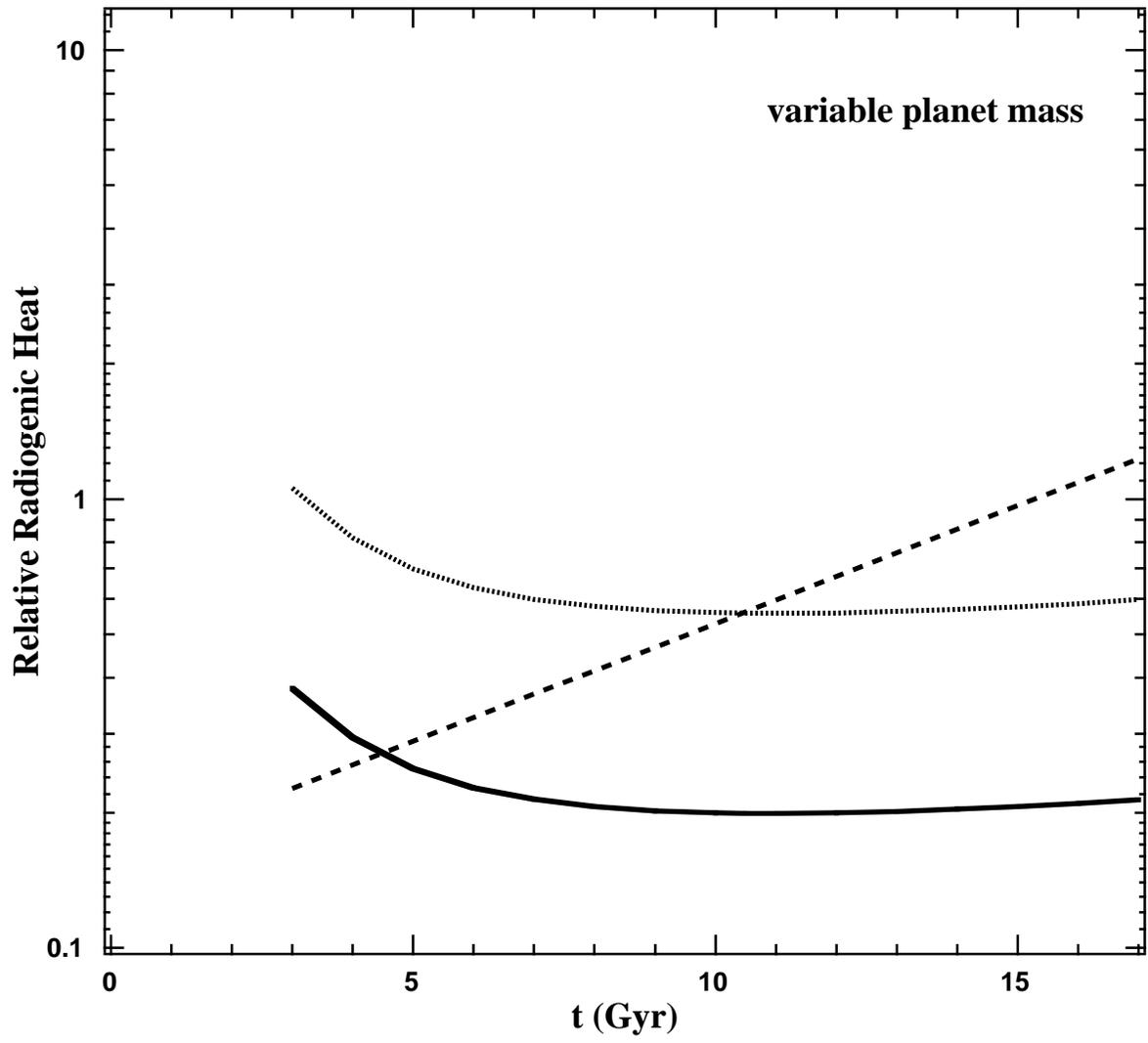
\end{document}